\documentclass[nofootinbib]{revtex4} 
\usepackage[a4paper,left=1.5cm,right=1.5cm,top=3cm,bottom=3cm]{geometry}

\usepackage{amssymb,amsmath,amsfonts,mathrsfs}
\usepackage[dvipsnames]{xcolor}
\usepackage{graphicx}
\usepackage{longtable}
\usepackage{verbatim}
\usepackage{color}
\usepackage[mathscr]{euscript}
\usepackage{framed}
\usepackage{mdframed}

\usepackage{color}
\usepackage{hyperref}
\hypersetup{colorlinks, citecolor=bluscuro, linkcolor=black, urlcolor=bluscuro}
\definecolor{rossos}{cmyk}{0,1,1,0.55}
\definecolor{bluscuro}{rgb}{0.15, 0.2, .85}
\definecolor{bluchiaro}{cmyk}{1,.3,0.,0.1}

\graphicspath{{./Figures/}}

\newcommand{\calH}{\mathcal{H}}
\newcommand{\med}[1]{\langle #1\rangle}

\newcommand{\dd}{{\rm d}}

\newcommand{\be}{\begin{equation}}
\newcommand{\ee}{\end{equation}}
\newcommand{\bea}{\begin{eqnarray}}
\newcommand{\eea}{\end{eqnarray}}
\def\beq{\begin{equation}}
\def\eeq{\end{equation}}

\def\vp{\varphi}

\def\d{{\rm d}}

\def\vk{{\vec{k}}}

\renewcommand{\L}{\mathrm{L}}
\newcommand{\R}{\mathrm{R}}

\def\beqa{\begin{eqnarray}}

\def\eeqa{\end{eqnarray}}
\def\lsim{\mathrel{\rlap{\lower4pt\hbox{\hskip0.5pt$\sim$}}
 \raise1pt\hbox{$<$}}}         
\def\gsim{\mathrel{\rlap{\lower4pt\hbox{\hskip0.5pt$\sim$}}
 \raise1pt\hbox{$>$}}}         

\def\vp{\varphi}

\def\d{{\rm d}}
\def\vx{{\vec{x}}}

\def\vk{{\vec{k}}}

\def\vp{{\vec{p}}}

\def\d{{\rm d}}
\def\lisa{\text{\tiny LISA}}
\def\PBH{\text{\tiny PBH}}
\def\M{{\tiny M}}

\newcommand{\Ic}{\mathcal{I}_c}
\newcommand{\Is}{\mathcal{I}_s}

\newcommand{\Pz}{\mathcal P_\zeta}
\newcommand{\Ph}{\mathcal P_h}
\newcommand{\Bh}{S_h}

\renewcommand{\S}{\mathscr S}
\newcommand{\OGW}{\Omega_\text{GW}}
\newcommand{\rhoGW}{\rho_\text{GW}}
\newcommand{\etaf}{\eta_f}
\newcommand{\rE}{r_{\text{\tiny E}}}
\newcommand{\rS}{r_{\text{\tiny S}}}
\newcommand{\MPBH}{M_{\text{\tiny PBH}}}
\newcommand{\colboxed}[1]{#1}
\def\eeqa{\end{eqnarray}}
\def\la{~\mbox{\raisebox{-.6ex}{$\stackrel{<}{\sim}$}}~}

\def\bq{\begin{quote}}
\def\eq{\end{quote}}

 at 10truept
\newcommand{\arXiv}[2]{\href{http://arxiv.org/pdf/#1}{{\tt [#2/#1]}}}
\newcommand{\arXivold}[1]{\href{http://arxiv.org/pdf/#1}{{\tt [#1]}}}

\usepackage[normalem]{ulem}

\numberwithin{equation}{section}
\renewcommand\theequation{\arabic{section}.\arabic{equation}}

\begin{document}
\def\thefootnote{\fnsymbol{footnote}}

\begin{center}
{\Large \textbf{Testing   Primordial Black Holes as Dark Matter   through LISA }}\\
\vspace{0.5cm}
{\large N. Bartolo$^{\rm a,b,c}$, V. De Luca$^{\rm d}$, G. Franciolini$^{\rm d}$,
 M. Peloso$^{\rm a,b}$,
	 D. Racco$^{\rm d, e}$ and A.~Riotto$^{\rm d}$}
\\[0.5cm]
{\small
\textit{$^{\rm a}$
Dipartimento di Fisica e Astronomia ``G. Galilei",
Universit\`a degli Studi di Padova, via Marzolo 8, I-35131 Padova, Italy}}

{\small
	\textit{$^{\rm b}$
		INFN, Sezione di Padova,
		via Marzolo 8, I-35131 Padova, Italy}}

{\small
	\textit{$^{\rm c}$
		INAF - Osservatorio Astronomico di Padova, Vicolo dell'Osservatorio 5, I-35122 Padova, Italy}}

{\small
\textit{$^{\rm d}$Department of Theoretical Physics and Center for Astroparticle Physics (CAP) \\
24 quai E. Ansermet, CH-1211 Geneva 4, Switzerland}}


{\small
\textit{$^{\rm e}$Perimeter Institute for Theoretical Physics, 31 Caroline St.~N., Waterloo, Ontario N2L 2Y5, Canada
}}
\end{center}
\vspace{.7cm}

\noindent
\begin{center}
 {\small \textbf{Abstract}}
 \end{center}
{\small The idea that primordial black holes (PBHs) can comprise most of the dark matter of the universe has recently reacquired a lot of momentum.
 Observational constraints, however, rule out this possibility for most of the PBH masses, with a notable exception  around $10^{-12} M_\odot$.
 These light PBHs may be originated when a sizeable  comoving curvature perturbation generated during inflation re-enters the horizon during the radiation phase.
 During such a stage, it is unavoidable that gravitational waves (GWs) are generated. Since their source is quadratic in the curvature perturbations, these GWs are generated  fully non-Gaussian.
Their  frequency today is about the mHz, which is exactly the range where  the LISA mission has the maximum of its sensitivity. This is certainly an impressive coincidence.
We show that this scenario of PBHs as dark matter can be tested by LISA by measuring the GW two-point correlator. On the other hand, we show that the short observation time (as compared to the age of the universe) and propagation effects of the GWs across the perturbed universe from the production point to the LISA detector suppress the bispectrum to an unobservable level. This suppression is completely general and not specific to our model.}

\vspace{0.3cm}
\noindent
\def\thefootnote{\arabic{footnote}}
\setcounter{footnote}{0}

\noindent
\section{Introduction}
\noindent
The possible presence and composition of dark matter (DM) in our universe constitutes one of the open questions in physics \cite{bertone}. The first direct observation of GWs generated by the merging of two $\sim 30 M_\odot$ black holes \cite{ligo} has increased the attention to the possibility that all (or a significant fraction of) the dark matter is composed by PBHs (see Refs. \cite{Bird:2016dcv,Clesse:2016vqa,Sasaki:2016jop,juan} and \cite{revPBH,revPBH1} for  recent reviews).
Inflation and a mechanism to enhance the comoving curvature perturbation $\zeta$ \cite{s1,s2,s3} at scales smaller with respect to the CMB ones are the only ingredients needed by the simplest models describing the PBH formation without the request of any physics beyond the Standard Model. In fact, the perturbations  themselves of the Standard Model Higgs
may be responsible for the growing of the comoving curvature perturbations during inflation \cite{Espinosa:2017sgp}.

Perturbations generated during inflation are transferred  to radiation through  the reheating process after inflation.
After they re-enter the horizon, a region  collapses to  a PBH if  the density contrast  (during the  radiation era)
\be
\Delta(\vec x)=\frac{4}{9a^2H^2}\nabla^2\zeta(\vec x)
\ee
 is larger than the  critical value $\Delta_{\rm c}$ which  depends on the shape of the power spectrum \cite{musco}.
The temperature at which the collapse takes place is
\be
T_\M\simeq 10^{-1} \left(\frac{106.75}{g_*}\right)^{1/4} \left(\frac{M_\odot}{M}\right)^{1/2}\,{\rm GeV},
\ee
where we have normalized $g_*$ to the effective number of the Standard Model degrees of freedom.
We define the power spectrum of the comoving curvature perturbation as
\begin{equation}
	\big<\zeta(\vk_1)\zeta(\vk_2)\big>' =\frac{2\pi^2}{k_1^3} \Pz(k_1),
	\label{eq: def P zeta}
\end{equation}
where we used the prime to indicate the rescaled two point function without the $(2\pi)^3$ and the Dirac delta for the momentum conservation.

It is useful to define the variance of the density contrast as
\be
\sigma^2_\Delta(M) =\int_0^\infty {\rm d}\ln k \,W^2(k,R_H){\cal P}_{\Delta}(k),
\ee
where we inserted a Gaussian window function $W(k,R_H)$ to smooth out the density contrast on scales given by the comoving horizon $R_H\sim 1/aH$ and  the density contrast power spectrum
\be
{\cal P}_{\Delta}(k)=\left(\frac{4k^2}{9a^2H^2}\right)^2 \Pz(k).
\ee
The mass fraction $\beta_\M$ indicating the portion of energy density of the universe collapsed into PBHs at the time of formation is
\be
\label{beta}
\beta_\M=\int_{\Delta_{\rm c}}^\infty \frac{{\rm d}\Delta}{\sqrt{2\pi}\,\sigma_\Delta}e^{-\Delta^2/2\sigma_\Delta^2}\simeq\frac{\sigma_\Delta}{\Delta_{\rm c}\sqrt{2\pi}} e^{-\Delta_{\rm c}^2/2\sigma^2_\Delta}\, ,
\ee
under the assumption of a Gaussian probability distribution. For the non-Gaussian extension see \cite{ng}.
The present abundance of DM in the form of PBHs per logarithmic mass interval $\mathrm d \ln M$ is given by
\be
\label{f}
f_\PBH(M) \equiv  \frac{1}{\rho_{\text{\tiny DM}}} \frac{{\rm d} \, \rho_\PBH}{{\rm d} \ln M} \simeq\left(\frac{\beta_\M}{6\cdot 10^{-9}}\right)\left(\frac{\gamma}{0.2}\right)^{1/2}\left(\frac{106.75}{g_*}\right)^{1/4}\left(\frac{M_\odot}{M}\right)^{1/2},
\ee
where $\gamma<1$ is a parameter introduced to take into account the efficiency of the collapse and, for the masses of interest, the number of relativistic degrees of freedom $g_*$ can be taken to be the SM value 106.75.

A peculiar feature of such models is that, after being generated during the last stages of inflation, the sizeable curvature perturbations unavoidably behave as a (second-order) source \cite{Acquaviva:2002ud, Mollerach:2003nq, Ananda:2006af, Baumann:2007zm} of primordial GWs at horizon re-entry \cite{jap}.
One can relate the peak frequency of such GWs, close to the characteristic frequency of the corresponding curvature perturbations which collapse to form PBHs, to its mass $M$ by using entropy conservation
\be
\label{f}
M\simeq 33\,\gamma \left(\frac{10^{-9}\,{\rm Hz}}{f}\right)^2 M_\odot.
\ee
\begin{figure}[t!]
\includegraphics[width=0.65\columnwidth]{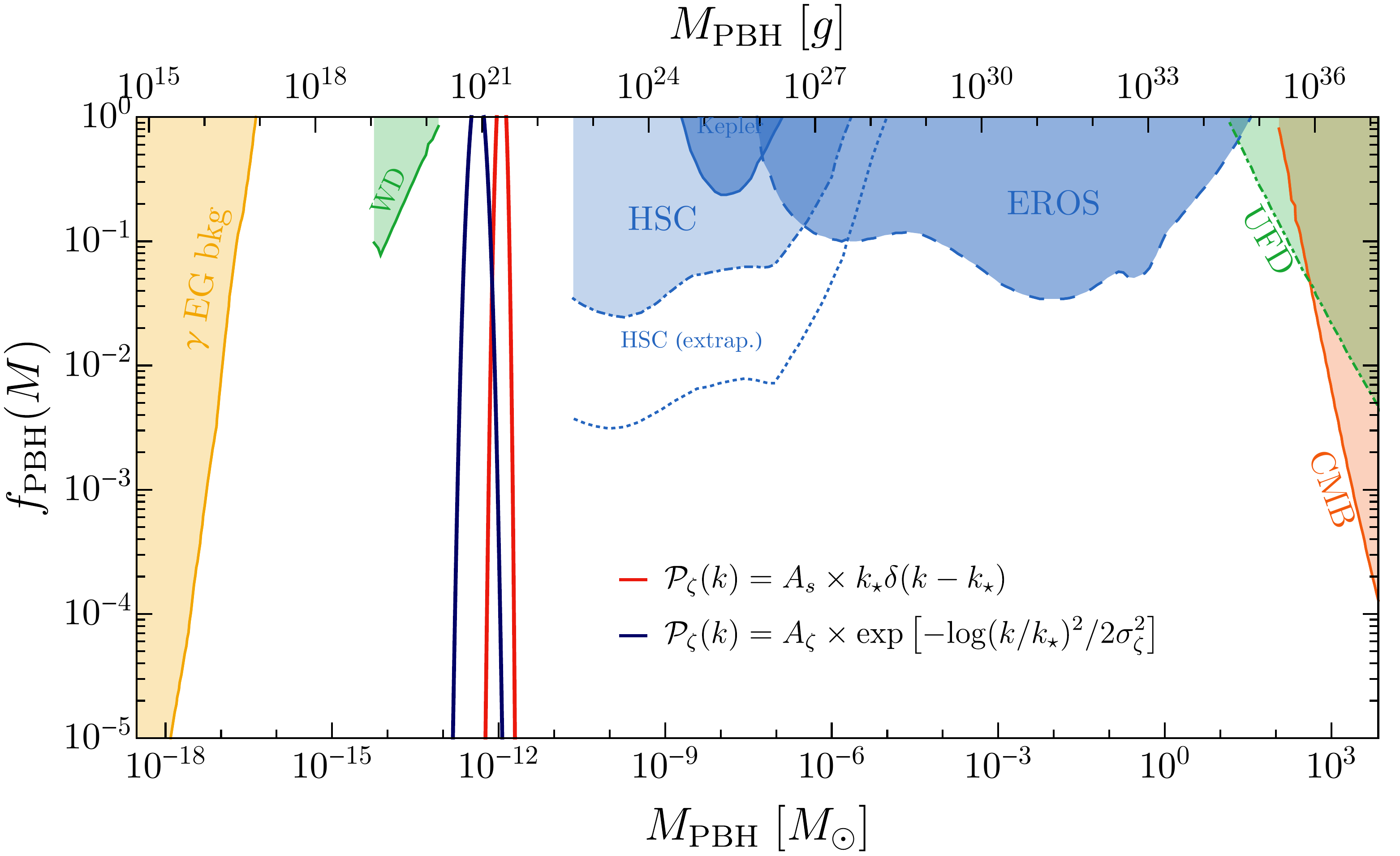}
\caption{
Overview on the present experimental constraints on the abundance of PBH for a monochromatic spectrum (from Ref. \cite{japanese} and references therein):
 in orange, constraints from the CMB; in green, dynamical constraints from White Dwarves and Ultra-Faint Dwarf galaxies; in blue, micro- and milli- lensing observations from Eros, Kepler, Subaru HSC;
 in yellow, the observations of extra-galactic $\gamma$-ray background.
Superimposed can be found the PBH abundances as a function of mass obtained for both power spectra in Eqs.~\eqref{Pz-delta} and \eqref{Pz-gauss}, where we have defined $k_\star=2\pi f_\lisa$. The total abundance is obtained by integrating over the masses and the parameters have been chosen to get a PBH abundance equal to the one of dark matter, respectively $A_s = 0.033$, $A_{\zeta} = 0.044$ and $\sigma_{\zeta} = 0.5$.
}
\label{fig: Omega PBH}
\end{figure}
\noindent
Choosing as frequency the one at which the Laser Interferometer Space Antenna (LISA) project  \cite{Lisa}  has its maximum sensitivity, i.e. $f_\lisa\simeq 3.4$ mHz,
Eq.~(\ref{f}) gives a mass $M\simeq  10^{-12} M_\odot$ (taking $\gamma\simeq 0.2$). Therefore, as also suggested in Ref.  \cite{gbp}, LISA measurements can provide useful information on PBH dark matter of such small masses.

As shown in Fig.~\ref{fig: Omega PBH}, the fact that the current observational constraints on the PBH abundances of such masses are missing \cite{k}, permitting $f_\PBH\simeq 1$, is a serendipity.
This is possible due to the fact that  the Subaru HSC  microlensing  \cite{hsc}  constraint  needs to be cut around the value $10^{-11} M_\odot$ under which  the geometric optics approximation is not valid for radiation in the optical wavelength \cite{hsc1,k}.
Another constraint analysed in the literature comes from the presence of neutron stars in globular clusters \cite{neut}, but we do not include it because it is based on controversial assumptions about the dark matter density in these systems.
We collect in Appendix~\ref{app:PBH constraints} a more detailed discussion of the issues related to these observational constraints.~\footnote{We briefly comment on the high-mass portion of Fig.~\ref{fig: Omega PBH}.
The Ultra-Faint Dwarf (UFD) galaxy constraint arises from the fact that PBHs of this mass would cause the dissolution of star clusters observed in UFDs such as Eridanus II \cite{Brandt:2016aco}; this constraint is strongly weakened in the presence of an intermediate-mass black hole, providing a binding energy that stabilizes the cluster \cite{Brandt:2016aco,Li:2016utv}. We thank Juan Garc\'ia-Bellido for discussions on this issue.
Secondly, we do not show in Fig.~\ref{fig: Omega PBH} the lensing bounds related to the measured luminosities of Supernov\ae\ Ia derived in \cite{Zumalacarregui:2017qqd,Garcia-Bellido:2017imq}, which constrain the abundance of PBHs above $1\,M_\odot$. We are not showing also the stronger bounds from CMB arising from disk-accretion \cite{disk-acc}. We also omitted the constraints coming from Lyman$-\alpha$ forest observations  \cite{murgia}, which overlap with the ones from UFD. 
Similarly, in the low-mass region, we do not show the constraints from \cite{cirelli} related the production of cosmic rays from evaporating PBHs, given that they  overlap with the constraints related to $\gamma-$rays produced by the PBHs evaporation.
}

It is certainly an exciting coincidence that the frequency range at which the LISA observatory has its maximum sensitivity corresponds to a region of the mass spectrum where the totality of dark matter composed of PBHs is allowed.

In this paper we demonstrate that, if dark matter is composed by PBHs of masses around $10^{-12} M_\odot$, then LISA will be able to measure  the power spectrum of the GWs necessarily generated by the production mechanism of PBHs. However, even though the GWs are non-Gaussian in nature, being sourced at second-order, their non-Gaussianity may not be measured by an experiment like LISA. The reason lies on the fact that the signal is  a superposition of waves coming  with momenta of different magnitudes and from all possible directions. Due to the relatively small observation time (as compared to the age of the universe) 
measurements at LISA cannot resolve modes of different momenta to a level of accuracy needed to preserve the coherency in the bispectrum. Moreover, even if we could construct a large array of LISA-like detectors, so to collect observations over a cosmological time, GWs coming from various directions propagate through different long wavelength density perturbations. This creates different time delays along different directions, thus making the initially correlated phases of the GWs fully uncorrelated. As we will show, this wipes out the bispectrum. It can also be seen as the central theorem in  action once the  ensemble averaging over the various directions is performed. Unfortunately, this effect seems to be general and not peculiar of our set-up and invalidates some results about the measurement of the tensor bispectrum through interferometers which appeared recently in the literature.

The reader should also be aware of the fact that the PBH abundance is exponentially sensitive to the amplitude of the variance. This means that a small
decrease of $\sigma_\Delta^2$ (and therefore the amplitude of the power spectrum of the comoving curvature perturbation) may reduce significantly the abundance. This, to some extent, plays in our favour as it implies that, even if  $f_\PBH\ll 1$, the corresponding GWs might be  anyway tested by LISA.

The paper is organised as follows.
In section II we describe the calculation leading to the GW power spectrum for two different shapes of the comoving curvature perturbations; section III is devoted to the calculation of the GW bispectrum.
 Section IV contains the details of the effects of the short observation time and of the propagation. Section V contains our conclusions.
The paper contains as well four   Appendices where some technicalities are provided, including an analysis of the LISA response functions for the bispectrum.

A short version of this paper presenting some of the main results can be found in Ref.~\cite{short}.

\section{The power spectrum of gravitational waves}
\noindent
The equation of motion for the GWs is found by expanding the tensor components of the Einstein's equations up to second-order in perturbations\footnote{We do not consider the free-streaming effect of neutrinos on the GW amplitude \cite{we}. }
\begin{equation}
	h_{ij}''+2\mathcal H h_{ij}'-\nabla^2 h_{ij}=-4 \mathcal T_{ij}{}^{ \ell m}\mathcal S_{\ell m},
	\label{eq: eom GW1}
\end{equation}
where we defined $'$ to denote the derivative with respect to conformal time $\eta$, $\dd\eta=\dd t/a$,  $\mathcal H=a'/a$ as the conformal Hubble parameter as a function of the scale factor $a(\eta)$ and the source term $\mathcal S_{\ell m}$ which, in a radiation dominated (RD) universe, takes the form
\cite{Acquaviva:2002ud}
\begin{equation}
	\label{psi}
	\mathcal S_{ij}=4\Psi\partial_i\partial_j\Psi+2\partial_i\Psi\partial_j\Psi-\partial_i\left(\frac{\Psi'}{\mathcal H}+\Psi\right)\partial_j\left(\frac{\Psi'}{\mathcal H}+\Psi\right).
\end{equation}
We note that  the mechanism of generation of GWs takes place when the relevant modes re-enter the Hubble horizon; in the case of interest, this happens deep into the radiation-dominated epoch.
It is also evident that the source is intrinsically second-order in the scalar perturbation $\Psi$.
For this reason the GWs generated are expected to feature an intrinsic non-Gaussian nature.
Additionally, since  the source contains two spatial derivatives, the resulting bispectrum in momentum space is expected to peak in the equilateral configuration.
The tensor $\mathcal T_{ij}{}^{\ell m}$ contracted with the source term  in Eq.~\eqref{eq: eom GW1} acts as a  projector selecting the  transverse and traceless components.
Its definition in Fourier space takes the form
\begin{equation}
 \tilde{\mathcal T}_{ij}{}^{\ell m}(\vk)=e_{ij}^{\rm L}(\vk)\otimes e^{{\rm L} \ell m}(\vk)
  + e_{ij}^{\rm R}(\vk)\otimes e^{{\rm R}\ell m}(\vk),
\end{equation}
where $e_{ij}^{\lambda}(\vk)$ are the polarisation tensors written in the chiral basis $(\L,\R)$.

The scalar perturbation $\Psi(\eta,\vk)$ appearing in Eq.~\eqref{psi} depends directly on the gauge invariant comoving curvature perturbation through the relation \cite{lrreview}
\begin{equation}
 \Psi(\eta,\vk)=\frac 23 T( k \eta ) \zeta(\vk) ,
 \label{eq: Psi to zeta}
\end{equation}
where the transfer function $T( k \eta )$ in the radiation-dominated  era is
\begin{equation}
  T(z)= \frac{9}{z^2}\left[ \frac{\sin (z/\sqrt 3)}{z/\sqrt 3} -\cos(z/\sqrt 3) \right].
\label{eq: transfer}
\end{equation}
By defining the dimensionless variables $x=p/k$ and $y=|\vec{k}-\vec{p}|/k$, the solution of the equation of motion \eqref{eq: eom GW1} can be recast in the following form
\begin{equation}
	\colboxed{
		h_\vk^{\lambda}(\eta) = \frac 49 \int \frac{\dd^3p}{(2\pi)^3} \frac{1}{k^3\eta}
		e^{\lambda}(\vk,\vp)  \zeta(\vp)\zeta(\vk-\vp) \Big[
		\Ic(x,y) \cos(k\eta) + \Is(x,y) \sin(k\eta)\Big],}
	\label{eq: h with Ic, Is}
\end{equation}
where we have introduced $e^{\lambda}(\vk,\vp) = e^{\lambda}_{ij}(\vk)p^i p^j$
and \cite{errgw}
\begin{equation}
	\begin{aligned}
		\Ic(x,y) &=4  \int_0^\infty \dd\tau \, \tau (-\sin \tau)  \Big[ 2T(x\tau)T(y\tau) + \Big(T(x\tau)
		+ x\tau\, T'(x\tau) \Big)\Big(T(y\tau) + y\tau\, T'(y\tau) \Big) \Big],
		\\
		\Is(x,y) &= 4 \int_0^\infty \dd\tau \, \tau (\cos \tau)  \Big\{ 2T(x\tau)T(y\tau) + \Big[T(x\tau) + x\tau\, T'(x\tau) \Big] \Big[T(y\tau) + y\tau\, T'(y\tau) \Big] \Big\}.
	\end{aligned}	\label{eq: Ic, Is}
\end{equation}
The complete analytical expressions of $\Ic(x,y)$ and $\Is(x,y)$ can be found in Appendix D of Ref. \cite{errgw} (see also Ref. \cite{Kohri:2018awv}).
We define the power spectrum of GWs using the same primed notation of Eq.~\eqref{eq: def P zeta} as
\begin{equation}
	\left\langle h^{\lambda_1}(\eta,\vk_1) h^{\lambda_2}(\eta,\vk_2)\right\rangle '\equiv
	\delta^{\lambda_1 \lambda_2}\,\frac{2\pi^2}{k_1^3} \Ph(\eta,k_1).
	\label{eq: def P h}
\end{equation}
After having computed the two-point function, in the radiation-dominated era we find $\Ph(\eta,k)$ to be
\begin{equation}
	\hspace{-1em}
	\colboxed{
		\Ph(\eta,k) =
		\frac{4}{81}\frac{1}{k^2\eta^2}
		\iint_\S \dd x\,\dd y
		\frac{x^2}{y^2} \left[1-\frac{(1+x^2-y^2)^2}{4x^2} \right]^2
		\Pz(kx) \Pz(ky)
		\big[ \cos^2(k\eta) \Ic^2
		+ \sin^2(k\eta) \Is^2
		+ \sin(2k\eta) \Ic\Is \big],
	}
	\label{eq: PS GWs}
\end{equation}
where $\S$ is the region in the $(x,y)$ plane allowed by the triangular inequality and shown in Fig.~2 of \cite{errgw}.
The power spectrum of GWs is directly connected to their energy density \cite{errgw}:
\begin{equation}
	\OGW(\eta,k) = \frac{\rhoGW(\eta,k)}{\rho_\text{cr}(\eta)}=\frac{1}{24} \left(\frac{k}{\calH(\eta)}\right)^2 \overline{\Ph(\eta,k)} ,
\end{equation}
where the overline  denotes an average over conformal time $\eta$.

So far we have assumed a radiation dominated universe with constant effective degrees of freedom  $g_*$ for the thermal radiation energy density (giving the analytic radiation-dominated solutions for $\eta$, $\calH$, etc.). In the Standard Model this will be approximately valid until some time $\eta_f$ before the top quarks start to annihilate and $g_*$ decreases.
The gravitational wave density scales $\propto 1/a^4$ because they are decoupled, but the radiation density $\rho_r(\eta_f)$ is related to the value today using conservation of entropy, giving
\begin{equation}
c_g \equiv \frac{a_f^4 \rho_r(\eta_f)}{\rho_r(\eta_0)} = \frac{g_*}{g_*^0} \left(\frac{g_{*S}^0}{g_{*S}}\right)^{4/3}
 \approx 0.4,
\end{equation}
where $g_{*S}$ is the effective degrees of freedom for entropy density and we assume $g_{*S}\approx g_* \approx 106.75$ from the Standard Model at time $\eta_f$.
Assuming $\rho_c(\eta_f)\approx \rho_r(\eta_f)$ we can then express the present density of GWs as
\begin{equation}
	\OGW(\eta_0,k) =  \frac{a_f^4\rhoGW(\eta_f,k)}{\rho_r(\eta_0)}\Omega_{r,0} =
 c_g
	\frac{\Omega_{r,0}}{24} \frac{k^2}{\calH (\etaf)^2} \overline{\Ph(\etaf,k)},
	\label{eq: Omega GW}
\end{equation}
and $\Omega_{r,0}$ is the present radiation energy density fraction if the neutrinos were massless.
The time average of the oscillating terms in Eq.~\eqref{eq: PS GWs}, together with the simplification of the time factor coming from $\mathcal{H}^2 (\eta_f) =1/\eta_f^2$ (valid before $\eta_f$), give the
current abundance of GWs
\begin{multline}
\OGW (\eta_0,k)= c_g \frac{ \Omega_{r,0}}{72} 
	\int_{-\frac{1}{\sqrt{3}}}^{\frac{1}{\sqrt{3}}}\dd d \int_{\frac{1}{\sqrt{3}}}^{\infty}\hspace{-5pt}\dd s
	\left[ \frac{(d^2-1/3)(s^2-1/3)}{s^2-d^2} \right]^2 \Pz\left(\frac{k\sqrt{3}}{2}(s+d)\right) \Pz\left(\frac{k\sqrt{3}}{2}(s-d)\right)
	\\
\times
	\left[ \Ic^2(x(d,s),y(d,s)) + \Is^2(x(d,s),y(d,s)) \right].
	\label{eq: Omega GW with PS}
\end{multline}
In the last step we also have redefined the integration variables as $d=(x-y)/\sqrt{3}$ and $s=(x+y)/\sqrt{3}$.

\subsection{The case of a Dirac delta power spectrum }
\noindent
In this subsection we make the idealized assumption that the scalar power spectrum has support in a single point
\begin{equation}
{\cal P}_\zeta (k) = A_s \, k_\star \delta \left( k - k_\star \right),
\label{Pz-delta}
\end{equation}
which can (obviously) be understood as the Gaussian case, discussed in the following subsection, in the limit of very small width.
This idealisation has the advantage that it allows to obtain exact analytic results for the amount of GWs produced at second-order by the scalar perturbations. In this subsection we compute the GW abundance, postponing the computation of the bispectrum to the next section.

Inserting Eq.~(\ref{Pz-delta}) in the expression (\ref{eq: Omega GW with PS}), the two Dirac delta functions allow to perform the two integrals, and  one obtains (see also Refs. \cite{saito,e})
\begin{equation}
\OGW(\eta_0,k) = \frac{c_g}{15552} \Omega_{r,0} A_s^2  \,  \frac{k^2}{k_\star^2} \left( \frac{4 k_\star^2}{k^2} -  1 \right)^2 \theta\left( 2 k_\star  - k \right) \left[ \Ic^2 \left( \frac{k_\star}{k} ,\,  \frac{k_\star}{k}  \right) + \Is^2 \left(  \frac{k_\star}{k} ,\,  \frac{k_\star}{k}  \right) \right].
\label{OGW-delta}
\end{equation}
This result is shown as a red line in Fig.~\ref{fig: Omega GW}.

\subsection{The case of a Gaussian power spectrum }
\begin{figure}[t!]
\includegraphics[width=0.6\columnwidth]{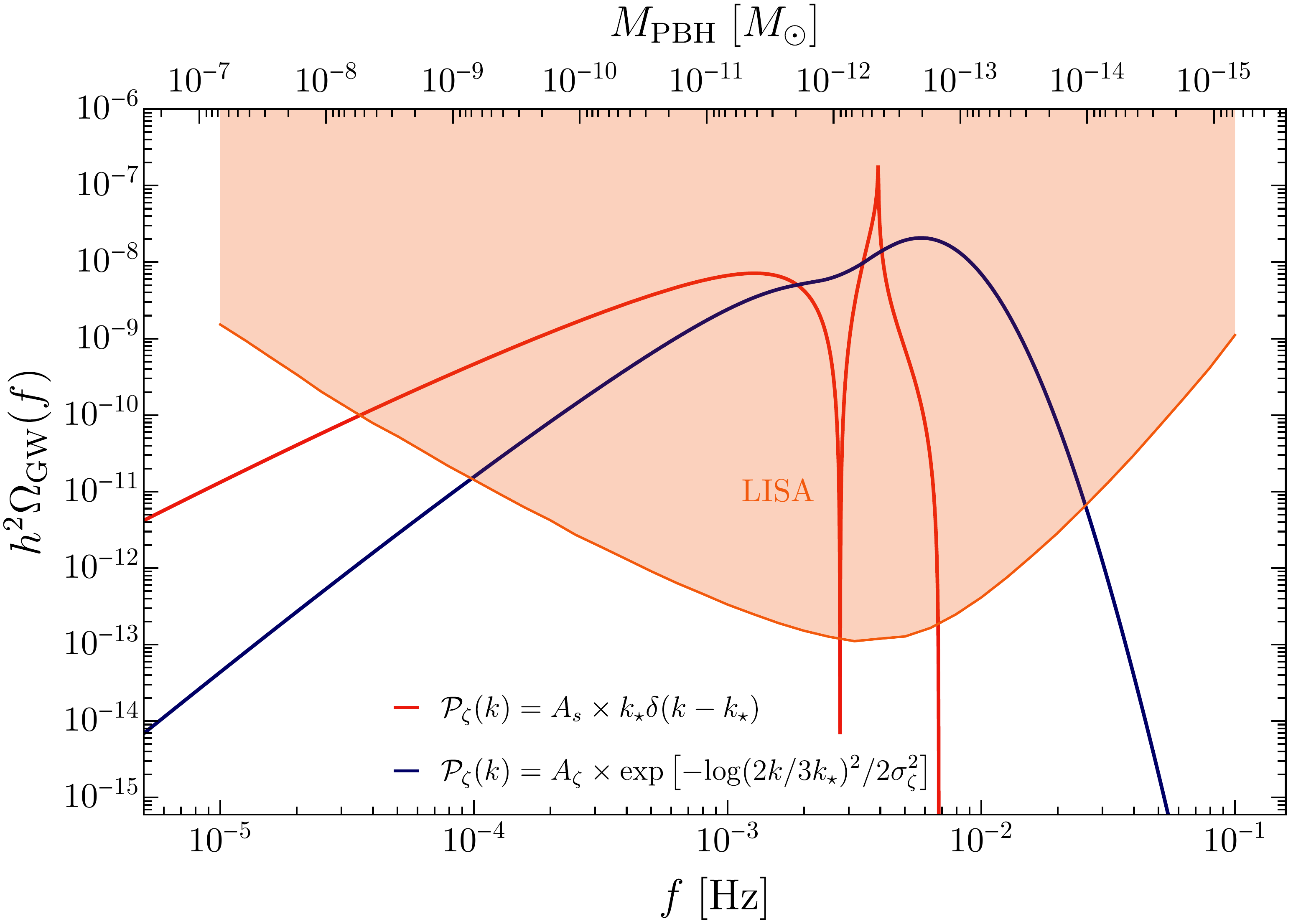}
\caption{
Comparison of the estimated sensitivity for LISA \cite{Audley:2017drz} (the proposed design (4y, 2.5 Gm of length, 6 links) is expected to yield a sensitivity in between the ones dubbed C1 and C2 in  Ref. \cite{Caprini:2015zlo}) with the GW abundance generated at second-order by the formation mechanism of PBHs for both power spectra in Eqs.~\eqref{Pz-delta} and \eqref{Pz-gauss}, where we used the following values for the parameters: $A_s = 0.033$, $A_{\zeta} = 0.044$ and $\sigma_{\zeta} = 0.5$.
 In the case of the monochromatic power spectrum, a resonant effect at $f \sim 2 f_{\rm LISA} / \sqrt{3}$ produces the spike, see for example Ref. \cite{Ananda:2006af}.
The slow fall-off  at low frequencies is an unphysical effect of assuming such a power spectrum, while physical spectra would typically give a white-noise ($\propto f^3$) behaviour \cite{errgw}, as one can observe in the second case, but with similar overall amplitudes.
}
\label{fig: Omega GW}
\end{figure}

\noindent
In this section we generalise the computation of the GWs energy density to the case of a Gaussian-like comoving curvature power spectrum.
We take the perturbation, enhanced with respect to the power spectrum on large CMB scales, to be
\be
\label{Pz-gauss}
{\cal P}^{g}_\zeta (k)=A_\zeta\,{\rm exp}\left(-\frac{\ln^2( 2 k/3k_\star)}{2\sigma_\zeta^2}\right).
\ee
This case differs from the former due to the wider shape of the power spectrum.
From the relation (\ref{f}) we can infer that PBHs of mass $\sim 10^{-12} M_{\odot}$ can amount for the totality of the dark matter if $\beta_{\M} \sim 6 \cdot 10^{-15}$ (considering $\gamma \sim 0.2 $), where
we assumed $k_\star R_H\simeq 1$ and the threshold to be $\Delta_{\rm c}\simeq 0.45$.
Its rigorous value is determined also by the shape of the power spectrum \cite{musco}  but, since the most relevant parameter $A_\zeta$ is not altered copiously, its impact on GWs is rather small.
The corresponding abundance of PBHs is shown by the blue line in Fig. \ref{fig: Omega PBH}, together with the current experimental bounds. We choose $A_\zeta\sim 0.044$ and $\sigma_\zeta =0.5$.
We stress again that a PBH of mass $\sim 10^{-12} M_\odot$ is associated to a scale $k_\star\sim k_\lisa=2\pi f_\lisa\simeq 2\cdot 10^{12}\,{\rm Mpc}^{-1}$.
The present abundance of GWs is given in Fig.~\ref{fig: Omega GW} where we see that it falls well within the sensitivity curves of LISA.
The different spectral shape with respect to the Dirac delta case is due to the tails of the Gaussian power spectrum of Eq.~\eqref{Pz-gauss}. At high frequencies, there is no upper bound at $2k_\star$ as in the Dirac delta case, because the scalar power spectrum is non vanishing for $k>k_\star$.
At lower frequencies, the spectral tilt for the Dirac-delta case is smaller than +2, whereas for the Gaussian case one can show that the tilt is about $\gtrsim 3$ by arguments similar to the ones exposed in  Ref. \cite{errgw}.
\newline
It is  clear that if PBHs of such masses form the totality (or a fraction of) dark matter, LISA will be able to measure the GWs sourced during the PBH formation time.

\section{The primordial bispectrum of GWs}
\noindent
We already stressed the fact that, being intrinsically at second order, the GWs are non-Gaussian; hence their primordial three-point correlator is not vanishing.
One can compute it following the procedure highlighted in Ref.~\cite{errgw}.
Computing the three-point function using Eq. \eqref{eq: h with Ic, Is} gives
\begin{eqnarray}
{\cal B}_{\lambda_i} \left(\vec{k}_i\right)&=&\Big< h_{\lambda_1}(\eta_1,\vk_1) h_{\lambda_2}(\eta_2,\vk_2) h_{\lambda_3}(\eta_3,\vk_3) \Big>' \nonumber \\
	&=& \left(\frac{8\pi}{9} \right)^3
	\int \dd^3 p_1
	\frac{1}{k_1^3 k_2^3 k_3^3 \eta_1 \eta_2\eta_3}
	\cdot
	e^*_{\lambda_1}(\vk_1,\vp_1) e^*_{\lambda_2}(\vk_2,\vp_2) e^*_{\lambda_3}(\vk_3,\vp_3)
	\frac{\Pz(p_1)}{p_1^3} \frac{\Pz(p_2)}{p_2^3} \frac{\Pz(p_3)}{p_3^3}\nonumber \\
	&\cdot&
	\left[ \Big(\cos(k_1\eta_1) \Ic\Big( \frac{p_1}{k_1},\frac{p_2}{k_1}\Big) + \sin(k_1\eta_1) \Is\Big( \frac{p_1}{k_1},\frac{p_2}{k_1}\Big)\Big)
	\cdot(1\rightarrow 2\, {\rm and}\, 2\rightarrow 3)\cdot  (1\rightarrow 3\, {\rm and}\, 2\rightarrow 1)\right],
	\label{bispectrum}
\end{eqnarray}
where $\vec{p}_2 = \vec{p}_1 - \vec{k}_1$ and $\vec{p}_3 = \vec{p}_1 + \vec{k}_3$.
\noindent

\subsection{The case of a  Dirac delta power spectrum }
\noindent
Inserting Eq.~(\ref{Pz-delta}) into Eq.~(\ref{bispectrum}), we obtain
\begin{eqnarray}
&&\left\langle h_{\lambda_1} \left( \eta_1,\vk_1 \right) h_{\lambda_2} \left( \eta_2,\vk_2 \right) h_{\lambda_3} \left( \eta_3,\vk_3 \right) \right\rangle' =
\left(\frac{8\pi}{9} \right)^3 \, \frac{A_s^3 \, k_\star^3}{k_1^3 k_2^3 k_3^3 \eta_1 \eta_2\eta_3} \,
\int \dd^3 p_1 \,
e_{\lambda_1}^* \left( \vk_1,\vp_1 \right) e_{\lambda_2}^* \left( \vk_2, \vp_1 - \vk_1 \right) e_{\lambda_3}^* \left( \vk_3 , \vp_1 + \vk_3 \right) \,
\nonumber\\
& & \quad\quad\quad \cdot \frac{\delta \left( p_1 - k_\star \right)}{k_\star^3}  \frac{\delta \left( \left\vert  \vec{p}_1 - \vec{k}_1 \right\vert - k_\star \right)}{k_\star^3}
\frac{\delta \left( \left\vert  \vec{p}_1 + \vec{k}_3 \right\vert - k_\star \right)}{k_\star^3} \,  \prod_{i=1}^3 \left[  \cos \left( k_i \eta_i \right) \Ic \left( \frac{k_\star}{k_i},\frac{k_\star}{k_i} \right) + \sin \left( k_i \eta_i \right) \Is \left( \frac{k_\star}{k_i},\frac{k_\star}{k_i} \right) \right]  .
\end{eqnarray}
As studied in Ref. \cite{Lisa}, the bispectrum depends on the orientation of the three vectors $\vec{k}_i$, as well as their magnitude. For definiteness, we fix
\begin{equation}
\vec{k}_1 = k_1 \, {\hat v}_1 ,\;\;\;\; \vec{k}_2 = k_2 \, {\hat v}_2 ,\;\;\;\;  \vec{k}_3 = - \vec{k}_1 - \vec{k}_2,
\label{3frame}
\end{equation}
where
\begin{eqnarray}
{\hat v}_1 = \left( \begin{array}{c} 1 \\ 0 \\ 0 \end{array} \right) ,\;\;
{\hat v}_2 = \left( \begin{array}{c} \frac{k_3^2-k_1^2-k_2^2}{2 k_1 k_2} \\  \sqrt{1-\left( \frac{k_3^2-k_1^2-k_2^2}{2 k_1
k_2} \right)^2} \\ 0  \end{array} \right) .
\label{v1hat-v2hat}
\end{eqnarray}
We then use  spherical coordinates for the integration vector $\vec{p}_1 = p_1
\left( \cos \theta ,\, \sin \theta \cos \phi ,\,  \sin \theta \sin \phi  \right)$, with $\cos \theta \equiv \xi$. We obtain (exploiting also the orthogonality of the
polarization operator $e^{\lambda}$)
\begin{eqnarray}
&&\left\langle h_{\lambda_1} \left( \eta_1,\vk_1 \right) h_{\lambda_2} \left( \eta_2,\vk_2 \right) h_{\lambda_3} \left( \eta_3,\vk_3 \right) \right\rangle' =
\left(\frac{8\pi}{9} \right)^3 \, \frac{A_s^3 }{k_\star^6 k_1^3 k_2^3 k_3^3 \eta_1 \eta_2\eta_3} \,
\prod_{i=1}^3 \left[  \cos \left( k_i \eta_i \right) \Ic \left( \frac{k_\star}{k_i},\frac{k_\star}{k_i} \right) + \sin \left( k_i \eta_i \right) \Is \left( \frac{k_\star}{k_i},\frac{k_\star}{k_i} \right) \right] \nonumber\\
&& \int_0^{\infty} \dd p_1 \, p_1^2 \, \int_{-1}^1 \dd \xi  \, \int_0^{2 \pi} \dd \phi \,
e_{\lambda_1}^* \left( \vk_1,\vp_1 \right) e_{\lambda_2}^* \left( \vk_2, \vp_1 - \vk_1 \right) e_{\lambda_3}^* \left( \vk_3 , \vp_1  \right)
\delta \left(p_1 - k_\star \right) \;
\delta \left( \sqrt{k_1^2 + p_1^2 - 2  k_1 p_1 \xi} - k_\star \right) \nonumber\\
& & \quad\quad\quad\quad \quad\quad \quad\quad
\delta \left( \left[ k_3^2 + p_1^2  + \frac{\left( - k_1^2 + k_2^2 - k_3^2 \right) p_1 \xi}{k_1} - \sqrt{2 \left( k_2^2+k_3^2 \right)- k_1^2 - \frac{\left( k_2^2 - k_3^2 \right)^2}{k_1^2}} p_1 \sqrt{1-\xi^2} \cos \phi \right]^{1/2} - k_\star \right) .
\nonumber\\
\end{eqnarray}
A careful study of the Dirac delta functions shows that the integral has support at the two points
\begin{equation}
\left( \vec{p}_1 \right)_{I,II} = k_1 \left( \frac{1}{2 }  ,\,
 \frac{-k_1^2+k_2^2+k_3^2}{8  \,  {\cal A} \left[ k_1 ,\, k_2 ,\, k_3 \right]},\, \pm \frac{\sqrt{16 {\cal A}^2 \left[ k_1 ,\, k_2 ,\, k_3 \right] k_\star^2 - k_1^2 k_2^2 k_3^2}}{4  {\cal A} \left[ k_1 ,\, k_2 ,\, k_3 \right]  k_1 }
 \right) \equiv k_1 \, \vec{q}_{I,II} ,
\label{p1-2}
\end{equation}
where
\begin{equation}
{\cal A} \left[ k_1 ,\, k_2 ,\, k_3 \right] \equiv  \frac{1}{4} \sqrt{\left( k_1+k_2+k_3 \right) \left( -k_1+k_2+k_3 \right)  \left( k_1-k_2+k_3 \right)   \left( k_1+k_2-k_3 \right) }
\end{equation}
is the area of the triangle of sides $k_i$. The support is present provided that
\be
 {\cal A} \left[ k_1 ,\, k_2 ,\, k_3 \right] > \frac{k_1 \, k_2 \, k_3}{4 \, k_\star}.
 \ee
With this in mind, the above integration gives (after some algebra)
\begin{multline}
{\cal B}_{\lambda_i} \left( \eta_i ,\, \vec{k}_i \right)  =  \frac{A_s^3 \theta \left(  {\cal A} \left[ r_1 ,\, r_2 ,\, r_3 \right] - \frac{r_1 \, r_2 \, r_3}{4 } \right)  }{k_1^2 k_2^2 k_3^2 \; k_\star^3 \, \eta_1 \eta_2 \eta_3}
 \frac{1024 \pi^3}{729}  {\cal D}_{\lambda_i } \left( {\hat k}_i ,\, r_i \right)  \\
 \cdot \left( \frac{16  \, {\cal A}^2 \left[ r_1 ,\, r_2 ,\, r_3 \right]}{r_1^2 r_2^2 r_3^2} - 1 \right)^{-1/2}
\frac{r_1^4}{r_2^2 \, r_3^2}  \prod_{i=1}^3
\left[ \frac{\mathcal{I}_i^*}{2} \; {\rm e}^{i\eta_i k_i} +  \frac{\mathcal{I}_i}{2} \; {\rm e}^{-i\eta_i k_i} \right],
\label{BS-exact}
\end{multline}
where we defined $r_i \equiv k_i/k_\star$, and introduced the combinations
\begin{equation}
\mathcal{I}_i \equiv \mathcal{I}\left(\frac{1}{r_i}\right) \equiv {\cal I}_c \left( \frac{1}{r_i} ,\, \frac{1}{r_i} \right) + i \, {\cal I}_s \left( \frac{1}{r_i} ,\, \frac{1}{r_i} \right),
\label{I-def}
\end{equation}
as well as  the contractions
\begin{equation}
 {\cal D}_{\lambda_i} \left( {\hat k}_i ,\, r_i \right)  \equiv e_{ab,\lambda_1}^* \left( {\hat k}_1 \right) e_{cd,\lambda_2}^* \left( {\hat k}_2 \right) e_{ef,\lambda_3}^* \left( {\hat k}_3 \right)
\left\{ \left[ \vec{q}_a \, \vec{q}_b  \left( \vec{q} - {\hat k}_1 \right)_c   \left( \vec{q} - {\hat k}_1 \right)_d  \vec{q}_e \, \vec{q}_f   \right]_I + \left[ {\rm same} \right]_{II} \right\} ,
\end{equation}
where we sum over the two points (\ref{p1-2}). The sum is performed as outlined in Appendix \ref{app:calD}.
We find that the contractions (and, therefore, the full bispectrum) are invariant under parity, namely under the $\L \; \leftrightarrow \; \R$ interchange. The resulting expressions are rather lengthy in the general $r_1 \neq r_2 \neq r_3$ case.
In Appendix  \ref{app:calD} we provide the explicit expression for the isosceles case $r_1=r_2$.
In the equilateral case $r_1=r_2=r_3$ the equal time bispectrum reads
\begin{equation}
{\cal B}_{\lambda_i}^{{\rm EQ}} \left( \eta ,\, |\vec{k}_i| = k \right)  =  \frac{A_s^3}{k_\star^3 \eta^3} \cdot \frac{1}{k^6} \; \frac{1024 \, \pi^3}{729} \, \frac{\theta \left( \sqrt{3} \, k_\star - k \right) }{\sqrt{\frac{3 k_\star^2}{k^2} - 1}} \; \left |\frac{1}{\sqrt{2}} {\cal I} \left( \frac{k_\star}{k} \right) \right |^3 \; \cdot \;
\left\{ \begin{array}{l}
 \frac{365}{6912}  - \frac{61}{192} \, \frac{k_\star^2}{k^2}  + \frac{9}{16}  \, \frac{k_\star^4}{k^4}   - \frac{1}{4 } \, \frac{k_\star^6}{k^6}  \;\; {\rm for \;\; RRR,\; LLL} , \\ \\
\frac{\left[ -4 +  \left( k / k_\star \right)^2 \right]^2 \left[ - 12 + 5 \left( k / k_\star \right)^2 \right] }{768 \left( k / k_\star \right)^6}  ,\;\; {\rm otherwise} .
\end{array} \right.
\label{BS-equil}
\end{equation}
where we have averaged over the oscillations of the amplitude (as done for the power spectrum in Ref.~\cite{Espinosa:2017sgp}), which amounts in the replacement
\begin{equation}
{\cal I}_c \left( \frac{1}{r_i} ,\, \frac{1}{r_i} \right) \cos \left( \eta_i k_i \right) +
{\cal I}_s \left( \frac{1}{r_i} ,\, \frac{1}{r_i} \right) \sin \left( \eta_i k_i \right)
\rightarrow \sqrt{ \frac{1}{2}
{\cal I}_c^2 \left( \frac{1}{r_i} ,\, \frac{1}{r_i} \right) + \frac{1}{2}
{\cal I}_s^2 \left( \frac{1}{r_i} ,\, \frac{1}{r_i} \right) } \equiv \frac{1}{\sqrt{2}} \left |{\cal I} \left( \frac{k_\star}{k_i} \right) \right | .
\label{nophase}
\end{equation}

\begin{figure}[t!]
\includegraphics[width=.495\textwidth]{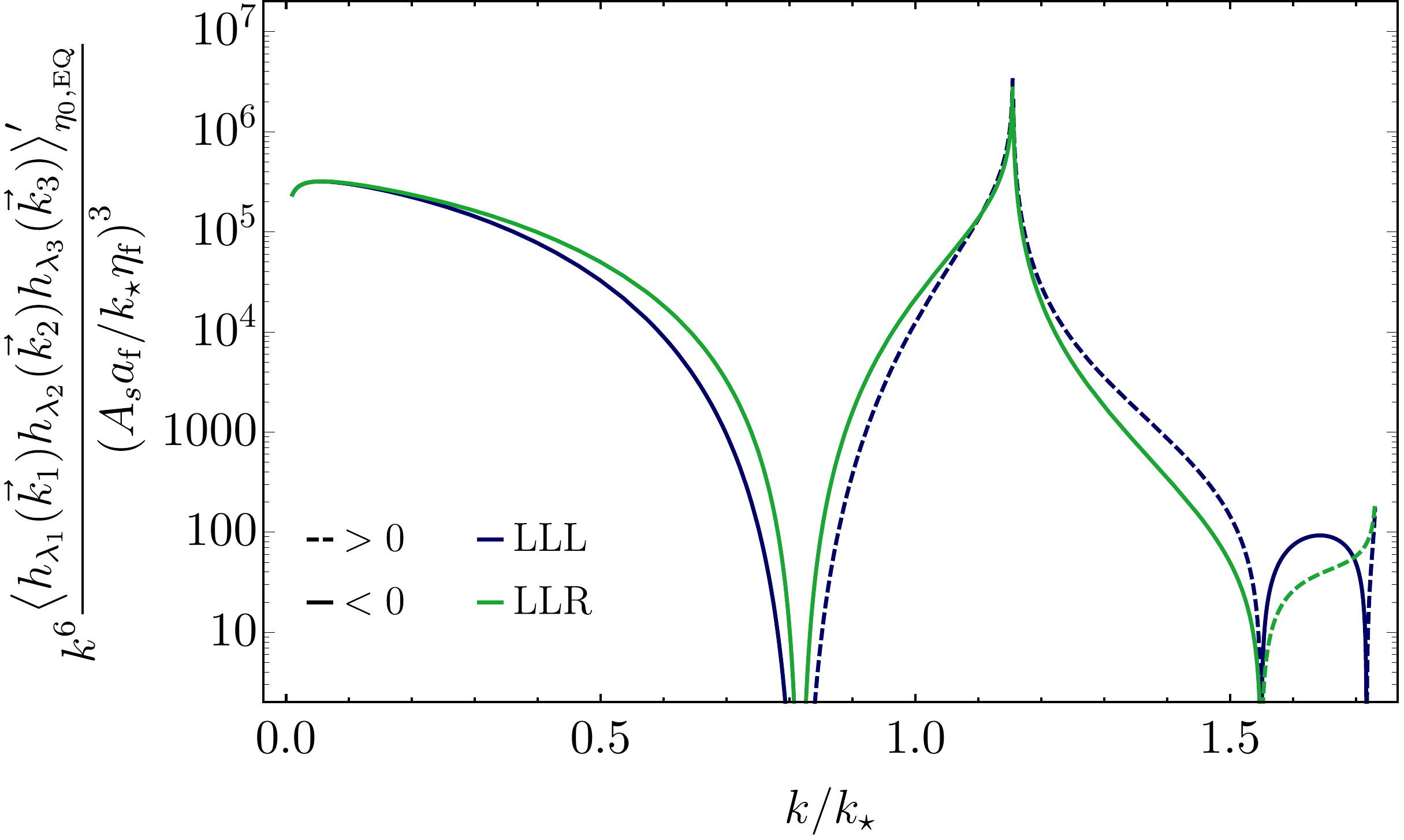}
\includegraphics[width=.495\textwidth]{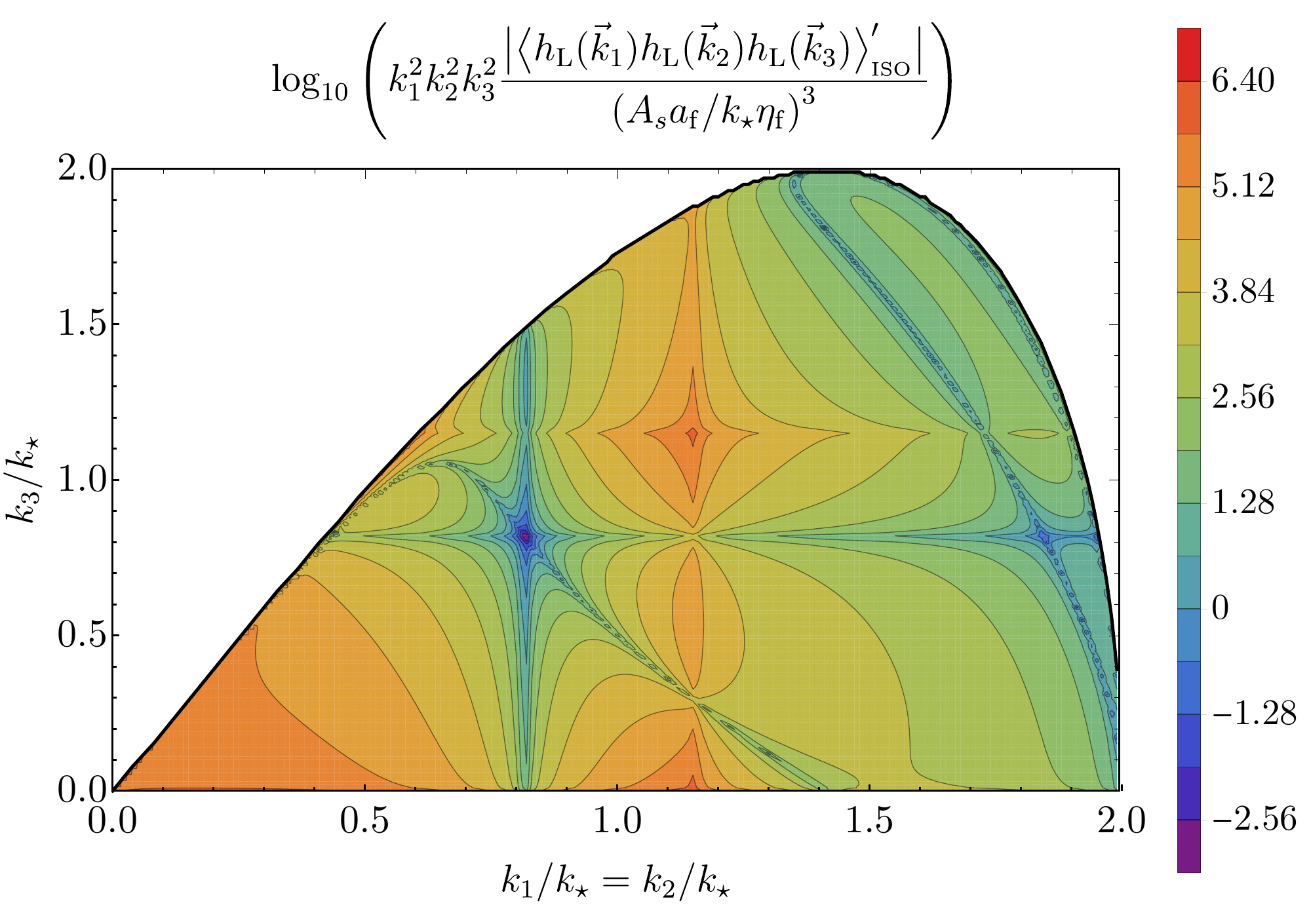}
\caption{{\bf Dirac delta power spectrum.}
 {\it Left:} Plot of the rescaled primordial bispectrum in the equilateral configuration. The bispectrum
vanishes in the outmost right part of the plot, namely for $k > \sqrt{3} k_{\star}$.  {\it Right:} Contour plot of the rescaled bispectrum for the isosceles case. The bispectrum vanishes in the white region.
}
\label{fig:BS}
\end{figure}
\noindent
In Fig.~\ref{fig:BS} we show the equilateral (left panel) and isosceles primordial bispectrum generated by the Dirac scalar power spectrum.
We see that the bispectrum is peaked in the equilateral configurations, at $k = 2 k_\star/\sqrt{3} $ (where the function $\cal{I}$ formally has a logarithmic divergence). This is clear also by looking at the plot of the shape, as defined in Eq.~\eqref{eq: shape},  shown in Fig.~\ref{fig:shapes} (left).

\subsection{The case of a Gaussian power spectrum}
\begin{figure}[t!]
\includegraphics[width=.483\textwidth]{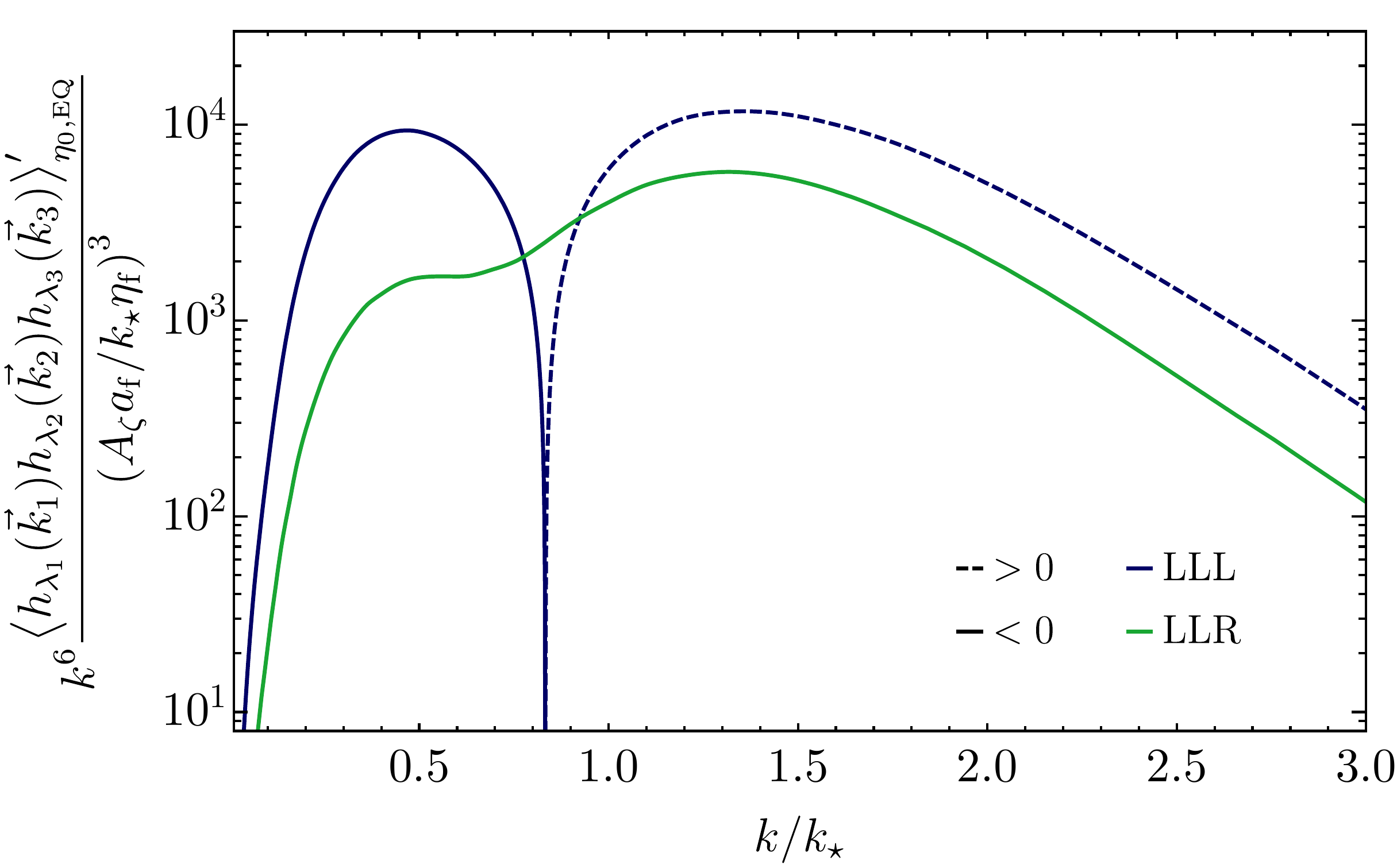}
\includegraphics[width=.507\textwidth]{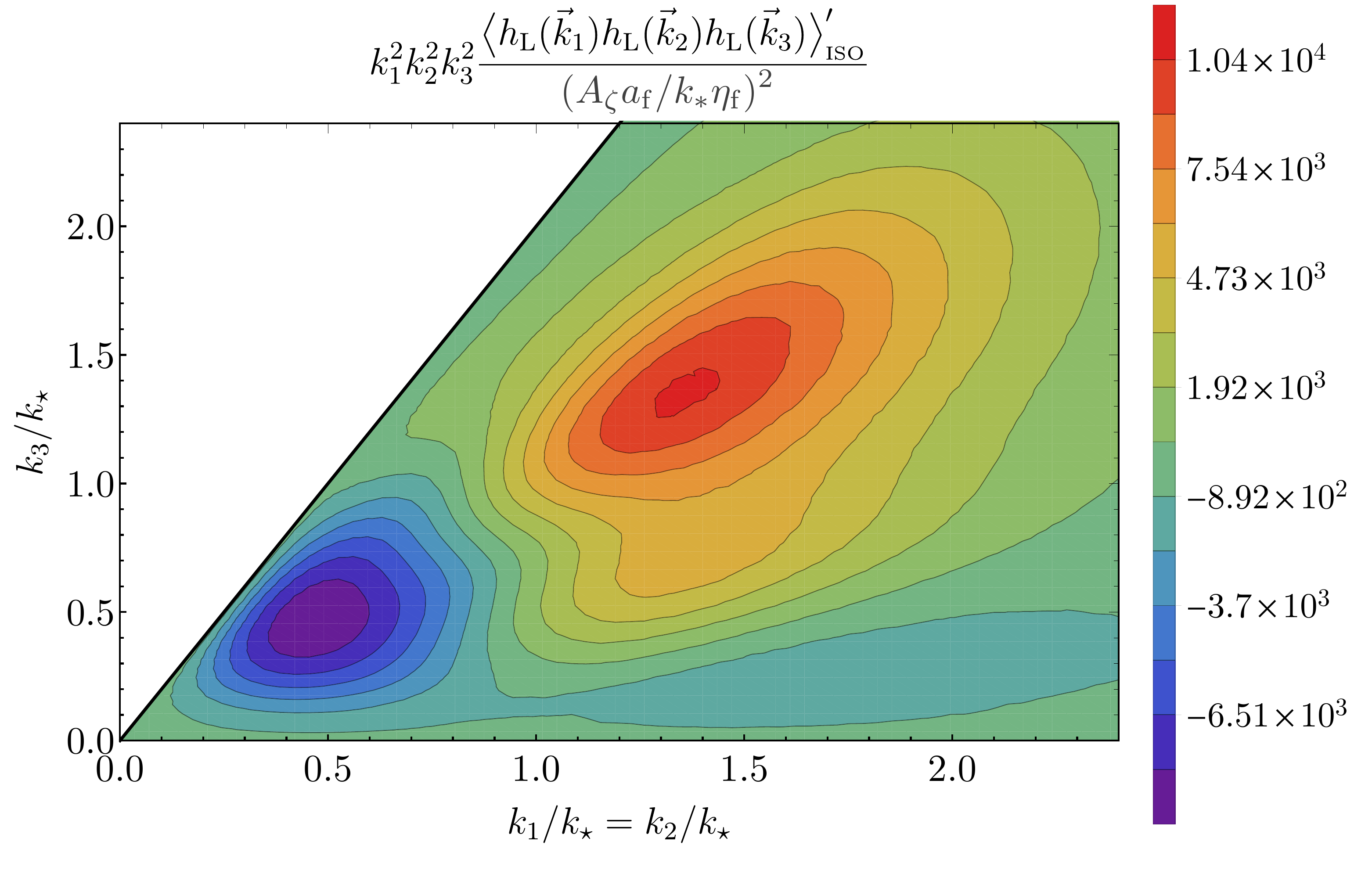}
\caption{{\bf Gaussian power spectrum.} {\it Left:} Plot of the rescaled bispectrum in the equilateral configuration.  {\it Right:} Contour plot of the rescaled bispectrum for the isosceles case. The white region is not allowed due to the triangular inequality imposed on $k_1$, $k_2$ and $k_3$ by conservation of momentum.
}
\label{fig:BSgauss}
\end{figure}

\noindent
It is now interesting to analyse the bispectrum corresponding to the Gaussian-like curvature perturbation power spectrum in Eq.~\eqref {Pz-gauss}.
First, we can compute it in the equilateral configuration, where we set $k_1 = k_2 = k_3$; the result is shown in the left panel of Fig. \ref{fig:BSgauss}.
A few comments can be made at this point. First, we see that the wider shape of the power spectrum compared to the Dirac delta one results in a lower peak in the equilateral configuration, making the two peaks with opposite sign in the LLL configuration comparable.
 Moreover, the width increase causes the bispectrum to be peaked at lower values of the momenta compared to $k_{\star}$. One has to keep in mind that, for sake of generality, in this case we assumed a power spectrum centred at a different momentum, namely $\sim3 k_{\star}/2$.
 As we shall see this change of the pivot scale, even though it does not introduce radical changes for what concern the PBH and the GWs abundances, can decrease the significance of the detection  at LISA.
 Finally, we note that the polarization configurations LLR=RRL (and their permutations) are suppressed with respect to the LLL=RRR ones.

In the right panel of Fig.~\ref{fig:BSgauss} one can see the behaviour of the rescaled three-point function of GWs in the isosceles configuration ($k_1=k_2$). Two important differences with respect to the Dirac case are its more regular profile and the absence of a cut-off present in the former case due to the Heaviside $\theta$-function.
\subsection{The shape of the primordial bispectrum}
\begin{figure}[t!]
\includegraphics[width=.49\textwidth]{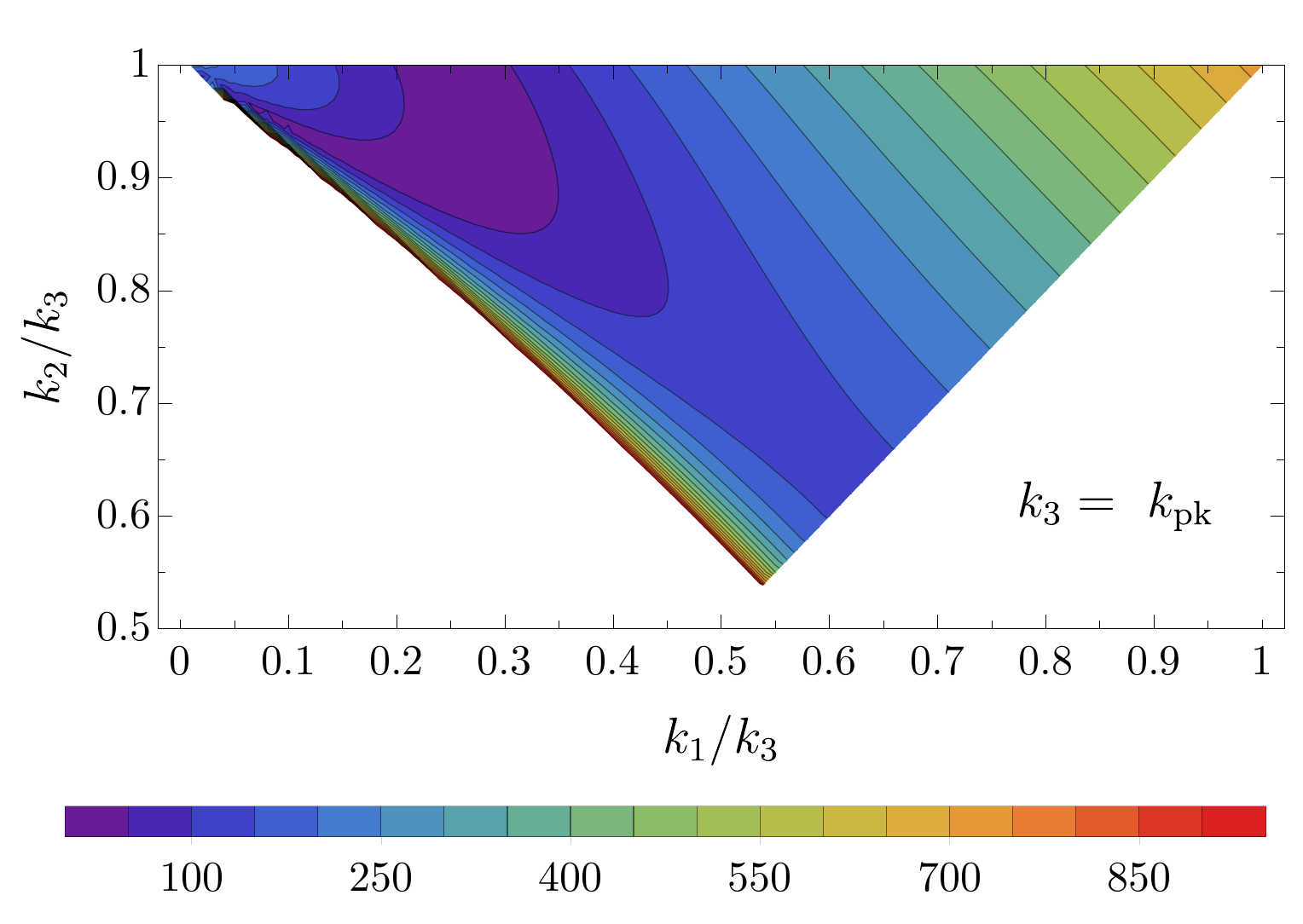}
\includegraphics[width=.49\textwidth]{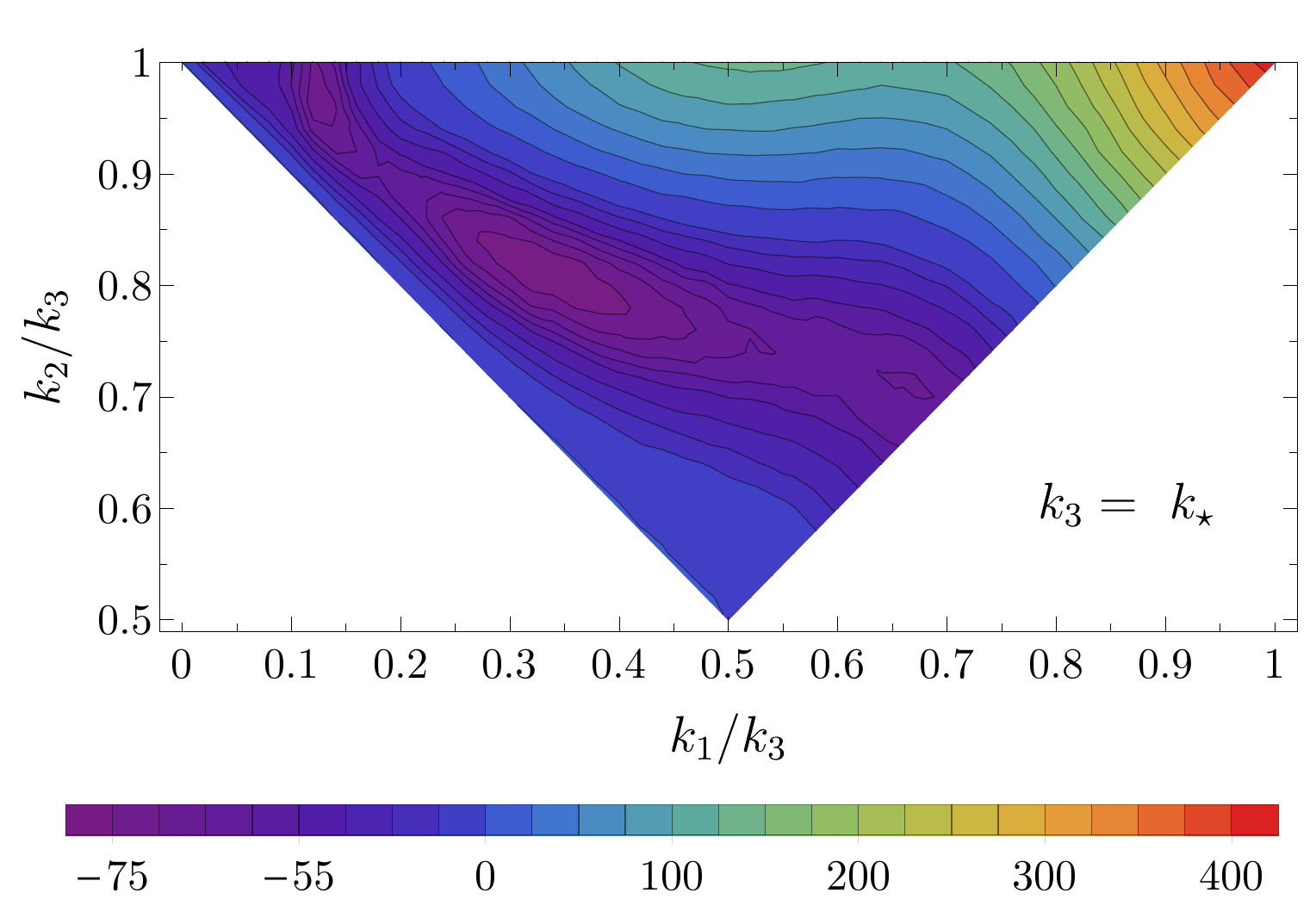}
\caption{{\bf Shape of the three-point function.} {\it Left:} Dirac delta case. {\it Right:} Gaussian case.
}
\label{fig:shapes}
\end{figure}

\noindent
We may  define the  shape for the bispectrum as
\begin{equation}
	\Bh^{\lambda_1\lambda_2\lambda_3}(\vk_1,\vk_2,\vk_3) = k_1^{2} k_2^{2} k_3^{2} \frac{\big< h_{\lambda_1}(\eta,\vk_1) h_{\lambda_2}(\eta,\vk_2) h_{\lambda_3}(\eta,\vk_3) \big>' }{\sqrt{\Ph(\eta,k_1)\Ph(\eta,k_2)\Ph(\eta,k_3)}}.
	\label{eq: shape}
\end{equation}
The shape, as defined in Eq.~\eqref{eq: shape}, is shown in Fig.~\ref{fig:shapes} (right).
Our findings  show  that the primordial bispectrum of GWs has its maximum at  the equilateral configuration, $k_1\simeq k_2\simeq k_3$. This comes about because the source of the GWs is composed by
gradients of the curvature perturbations when the latter
re-enters the horizon.   The measurement of this shape would by itself provide a consistency relation between the bispectrum and the power spectrum of GWs, which might help 
disentangle the signal from other possible sources.

\section{Measurement of the bispectrum}

The fact that the primordial bispectrum of GWs is sizeable does not automatically imply that it can be measured by LISA. Indeed, in this section we are going to show that the relatively short observation time (as compared to the age of the universe) and propagation of the GWs from the source points towards the detector across the perturbed universe largely suppresses the bispectrum signal. This point has not been noticed in the recent literature about the possibility of measuring the tensor bispectrum at interferometers. 

\subsection{The effect of the short observation time}

LISA can be considered as three time delay interferometers. Each interferometer is placed at the vertex of 
an (approximately) equilateral triangle, and the two arms of this interferometer connect the vertex with the other two vertices of the triangle. The various ``LISA channels'' are linear combinations of the time delays measured at the three vertices, as we discuss more in detail in Appendix \ref{app:LISA}. The time delay accumulated during a trip from the point $\vec{x}_1$ to the point $\vec{x}_2 = \vec{x}_1  + \vec{L}$, and return, is given by \cite{snr} 
\begin{eqnarray}
\Delta\eta \left( \eta_i \right) = \frac{L}{2} \int\frac{\d^3k}{\left( 2\pi \right)^3}\,e^{i \vec{k} \cdot\vec{x}_1} \sum_\lambda e_\lambda \left( \hat k,\hat L \right) \, \left[e^{ i k \eta_i}h_\lambda \left( \vec k \right) 
{\cal M} \left( \hat L\cdot\hat k,k \right) + e^{ -i k \eta_i}h^*_\lambda \left( -\vec k \right) {\cal M}^* \left( -\hat L\cdot\hat k,k \right) \right] \;, 
\label{Delta-eta} 
\end{eqnarray}
where $e_\lambda \left( \hat k,\hat L \right) \equiv e_{ab,\lambda} \left( \hat k \right) \, {\hat L}_a {\hat L}_b$, 
and 
\begin{equation}
{\cal M}(\hat L\cdot\hat k,k)=e^{i k L \left( 1-\hat L\cdot \hat k \right) / 2} \: {\rm sinc}\left[\frac{k L \left( 1-\hat L\cdot \hat k \right)}{2}\right] \;. 
\end{equation}

We are interested in the Fourier transform of the three point function of (\ref{Delta-eta}). This is proportional to the sum of quantities of the type 
\begin{eqnarray} 
& &\left\langle \Delta {\tilde \eta} \left( p_1 \right)  \Delta {\tilde \eta} \left( p_2 \right)  \Delta {\tilde \eta} \left( p_3 \right) \right\rangle \propto  \prod_{i=1}^{3} \int_{\eta_0-\delta \eta}^{\eta_0+\delta \eta} d \eta_i {\rm e}^{-i p_i \, \eta_i} \int \d^3 k_i \,e^{i\vec x_i \cdot \vec k_i}e^{+ i k_i \eta_i}{\cal M} \left( \hat L_i\cdot\hat k_i,k_i \right)  \, \left\langle h_{\lambda_1} \left( \vk_1 \right) h_{\lambda_2} \left( \vk_2 \right) h_{\lambda_3} \left( \vk_3 \right) \right\rangle \nonumber\\
& & \quad\quad 
 + \dots + \prod_{i=1}^{3} \int_{\eta_0-\delta \eta}^{\eta_0+\delta \eta} d \eta_i {\rm e}^{-i p_i \, \eta_i} \int \d^3 k_i \,e^{i\vec x_i \cdot \vec k_i}e^{- i k_i \eta_i}{\cal M}^* \left( - \hat L_i\cdot\hat k_i,k_i \right)  \, \left\langle h_{\lambda_1}^* \left( -\vk_1 \right) h_{\lambda_2}^* \left( -\vk_2 \right) h_{\lambda_3}^* \left( -\vk_3 \right) \right\rangle \,,
\end{eqnarray} 
where the observation is done at the present cosmological time, in the interval $\eta_0 - \delta \eta \leq \eta \leq \eta_0 + \delta \eta$. The integral over time in the first term results in 
\begin{equation}
\prod_{i=1}^{3} 2 \, \delta \eta
{\rm e}^{- i p_i \eta_0} \int \d^3 k_i \: 
{\rm sinc} \, \left[ \left( k_i - p_i \right) \delta \eta \right] {\rm e}^{i k_i \, \eta_0} \;  \,e^{i\vec x_i \cdot \vec k_i} 
{\cal M} \left( \hat L_i\cdot\hat k_i,k_i \right)  \, \left\langle h_{\lambda_1} \left( \vk_1 \right) h_{\lambda_2} \left( \vk_2 \right) h_{\lambda_3} \left( \vk_3 \right) \right\rangle \;, 
\label{measured}
\end{equation} 
and analogously for the other terms. Unless the observation time $\delta \eta$ is comparable to the age of the universe $\eta_0$ (see below) the phases ${\rm e}^{i (\pm k_i \eta_0)}$ are by far the most rapidly varying terms in the integrand of (\ref{measured})~\footnote{The presence of these phases was missed in Ref.~\cite{snr}, since the bispectrum considered there was a function only of the difference of the observation times of the three signals, and not of the average time $\eta_0$. This is a consequence of the ansatz for the bispectrum considered in that paper, for which the bispectrum resulted to be the sum of products of two unequal time power spectra. This ansatz does not appear to be consistent with the time evolution for the three GWs contributing to the bispectrum, which is encoded in (\ref{Delta-eta}).}. As a consequence, the integral over the magnitude of the momenta averages to (essentially) zero due to these fast oscillations, except from the configurations that satisfy $\eta_0 \sum_i \pm k_i \lsim 1$. We note that these phases do not enter in the time averaged power spectrum (\ref{eq: Omega GW}). Therefore, this suppression is not present for the observed power spectrum. 

The GW bispectrum vanishes unless the three momenta  satisfy $\vec{k}_1 +\vec{k}_2 + \vec{k}_3 = 0$. Therefore, given that the GW signal peaks at $k_1 \sim k_2 \sim k_3 = {\cal O } \left( k_* \right)$, and that $\eta_0 \, k_* \gg 1$, the above condition can be satisfied only if the three momenta are extremely well aligned. Assuming for example $\vec{k}_1 = k_1 \left( 1 ,\, 0 ,\, 0 \right)$ and $\vec{k}_2 = k_2 \left( \cos \theta ,\, \sin \theta ,\, 0 \right)$, with $\theta \ll 1$, we have $\eta_0 \left( k_1 + k_2 - k_3 \right) \simeq 
\frac{\eta_0 \, k_1 \, k_2 \, \theta^2}{2 \left( k_1 + k_2 \right) } = {\cal O } \left( \eta_0 \, k_* \, \theta^2 \right)$. 
For $k_* = {\cal O } \left( {\rm mHz} \right)$, corresponding to the peak of the LISA sensitivity, and for $\eta_0 \sim 10^{17} \, {\rm s}$, parametrically equal to the age of the universe, we must require $\theta \lsim 10^{-7}$. 
This corresponds to the negligible fraction $ \sim \left( \eta_0 \, k_* \right)^{-1} \ll 1 $ of the integration region in (\ref{measured}). This scaling is consistent with the fact that the GW signal reaching the detector at the time $\eta_0$ is generated in $N \propto \left( k_* \eta_0 \right)^2$ independent Hubble patches (on all possible line of sight directions). Then the scaling $1/\sqrt{N}$  is what one expects from the measurement of the bispectrum of the sum of $N$ independent signals. 

The phases in (\ref{measured}) cause the result to average to zero if the range of $k$ values included in each observed band is large compared to $1 / \eta_0$. The bandwidth that can in principle be resolved in the observation time $\delta \eta$ is $\sim 1/\delta \eta$, and this cancellation of the bispectrum disappears in the unphysical limit of $\delta \eta \sim \eta_0$. At the mathematical level, in the limit of $\delta \eta \to \infty$, the product $\delta \eta \; {\rm sinc} \, \left[ \left( k_i - p_i \right) \delta \eta \right] $ reduces to the Dirac $\delta-$function of argument $k_i - p_i$, forcing the momentum of the GW to be equal to the frequency of the corresponding time delay signal. This observation time cannot obviously be attained.

One might instead (at least, in principle) construct a measurement that effectively uses a large array of LISA-like detectors placed at large distances. These detectors could  capture the wavefront from each source horizon volume at multiple locations today, resulting in observable phase correlations. This would require having detectors that are spread over a cosmologically large volume, since the GWs generated by each region have 
spread over a shell of radius $\sim \eta_0$ today, and a sizeable portion of this volume must be sampled to obtain a significant correlation. Even in this case, however, the GW phases decorrelate, due to a different physical reason. Specifically, waves propagating in different directions experience a differently perturbed universe, and accumulate a different Shapiro time delay. This also reduces the bispectrum to an unobservable level, as we demonstrate in the next subsection.

\subsection{GW propagation in a perturbed universe}
\noindent

Even if we could collect measurements over a cosmologically long observation time, or from 
a cosmologically large region, the fact that GWs arriving to the detector propagate in a perturbed universe would by itself still render the bispectrum unobservable. The physical reason is the following. A non vanishing non-Gaussianity requires the correlation among the phases of the GWs at the detection
point. This is the reason why non-Gaussianity is a synonym of phase correlations in the same way Gaussianity is characterised by random phases \cite{Matsubara}. In the case studied in this paper, the phases of the GWs are correlated at the moment of production thanks to the coherence generated during the inflationary stage. However, this coherence is destroyed due to the fact that the GWs measured by LISA travel
from different directions. Because of the Shapiro time-delay, caused by the presence of large-scale gravitational potentials along the line of sight, different directions experience different delays, thus
destroying the phase correlation. We will show that this effect is not present in the power spectrum, but appears,  unfortunately, in the bispectrum under the form of a large suppression factor. This is  a consequence  of the central limit theorem: the  time delay does not   suppress  the non-Gaussianity of the signal arriving from a single line of sight. However, averaging over many directions makes the signal Gaussian.

We divide this discussion in three parts: in the first one we compute the solution for the GW propagation in the geometrical optic limit, in the second one we show that the power spectrum is unaffected, and in the third part we compute  the effect on the bispectrum.

\subsubsection{The propagation equation and its solution in the geometrical optic limit}
\noindent
Our starting point is the metric in the Newtonian longitudinal gauge (neglecting the shear such that $\Phi=\Psi$) and in cosmic time
\begin{equation}
g_{00} = - \left( 1 + 2 \Phi \right) \;\;,\;\;
g_{0i} = 0 \;\;,\;\;
g_{ij} =  a^2 \left[ \left( 1 - 2 \Phi \right) \delta_{ij} + h_{ij} \right].
\end{equation}
The spatial components of the Einstein tensor  up to second-order in the metric perturbations written above are
\begin{eqnarray}
G^i_j &=& \delta_{ij} \left( - H^2 - \frac{2 \ddot{a}}{a} \right) \nonumber\\
&& + \frac{1}{2} \ddot{h}_{ij} + \frac{3}{2} H \dot{h}_{ij} - \frac{1}{2 a^2} h_{ij,kk} + {\rm O } \left( \Phi \right) \nonumber\\
&& + \ \left( \ddot{\Phi} + 3 H \dot{\Phi} \right) h_{ij} - \frac{1}{a^2} \left[  \Phi h_{ij,kk} + \left( \Phi h_{ij} \right)_{,kk}
-  \left( \Phi_{,k} h_{ik} \right)_{,j} -  \left( \Phi_{,k} h_{kj} \right)_{,i} \right]  \nonumber\\
& & + \delta_{ij} \left[ - \frac{1}{a^2} h_{mn} \Phi_{,mn} \right] + {\cal O } \left( \Phi^2  \right) +  {\cal O } \left( h^2  \right).
\end{eqnarray}
In the following we study the propagation of the GWs in the matter-dominated phase neglecting the tiny generation of the GWs caused by the  $ {\cal O } \left( \Phi^2  \right) $ during the propagation.
From now on, we will work in the geometrical optic limit which amounts to assuming that the frequency of the GW is much larger than the typical momentum associated to the
gravitational potential $\Phi$.
To recover the suppression due to the time-delay, it is therefore enough to consider the leading term in an expansion in gradients of $\Phi$. The corresponding equation reads, going back to conformal time,
\begin{equation}
h_{ij}'' + 2 {\cal H} h_{ij}' -  \left( 1 + 4 \Phi \right) h_{ij,kk}  = 0,
\end{equation}
where the transverse-free and traceless part will be taken off by the projector operators later on.
In momentum space, we write the mode function as
\be
h_{ij}(\vec k,\eta)=h^A_{ij}(\vec k,\eta)+h^{*A}_{ij}(-\vec k,\eta),
\label{hAdef}
\ee
which makes the mode function real in coordinate space. We make the ansatz
\begin{eqnarray}
h^A_{ij} &=& A_{ij} {\rm e}^{i k \eta} {\rm e}^{i \int^\eta d \eta' F_A \left( \eta' \right)}.
\end{eqnarray}
If we  disregard the spatial derivatives acting on $F_{A}$ and   separate the equations at zero-th  order in $F_{A}$ and the rest, we obtain
\begin{eqnarray}
\left\{ \begin{array}{l}
A_{ij}'' + 2 i k A_{ij}' - k^2 A_{ij}
+ \frac{4}{\eta} \left[ A_{ij}' + i k A_{ij} \right] + k^2 A_{ij} = 0, \\
2 i F_A A_{ij}' + \left[ i F_A'  - 2 k F_A - F_A^2
+ \frac{4}{\eta} i F_A  + 4 k^2 \Phi \right] A_{ij} \simeq 0.
\end{array} \right.
\end{eqnarray}
The first  equation is  solved by
\begin{equation}
A_{ij} = \frac{C^A_{ij}}{k^2 \eta^2} \left( 1 + \frac{i}{k \eta} \right),
\end{equation}
with $C^{A}$ an  integration constant. We are interested in the sub-horizon limit of these solutions, where only the first term in the sums is kept. We then have $A'_{ij} = \frac{-2}{\eta} A_{ij}$. The second  equation then becomes
\begin{eqnarray}
i F_A'  - 2 k F_A - F_A^2   + 4 k^2 \Phi \simeq 0.
\end{eqnarray}
To leading order, we find $F_A  = 2 k \Phi$, such that
\begin{equation}
h^A_{ij} = A_{ij} {\rm e}^{i k \eta + 2 i k \int^\eta d \eta' \, \Phi \left( \eta' \right)}.
\end{equation}
To match with the solution (\ref{eq: h with Ic, Is}) we write $h^A_{ij}$, and its conjugate appearing in (\ref{hAdef}) as a sum of cosine and sine. Matching the wave function and its derivative in the sub-horizon limit gives
\begin{eqnarray}
h_\lambda \left( \vec{k} \right) &=& \frac{\eta_{\rm eq}}{\eta^2} \left[ h_\lambda^{RD,c} \left( \vec{k} \right) \,   \cos \left( \Omega_{\eta,\vec{k}}  \right)  + h_\lambda^{RD,s} \left( \vec{k} \right) \, \sin \left( \Omega_{\eta,\vec{k}}  \right)   \right] \nonumber\\
 & = & \frac{4}{9} \int \frac{d^3 p}{\left( 2 \pi \right)^3} \, \frac{\eta_{\rm eq}}{k^3 \eta^2} \, e_\lambda^* \left( \vec{k} ,\, \vec{p} \right) \zeta \left( \vec{p} \right) \zeta \left( \vec{k} - \vec{p} \right)
\left[ {\cal I}_c \left( x,y \right) \cos \left( \Omega_{\eta,\vec{k}}  \right) +  {\cal I}_s \left( x,y \right) \sin \left( \Omega_{\eta,\vec{k}}  \right) \right] 
\end{eqnarray}
where now
\begin{eqnarray}
\Omega_{\eta,\vec{k}} &=& k \, \eta  + 2 k \int_{\eta_{\rm eq}}^\eta d \eta' \, \Phi \left( \eta' ,\, \vec{x}_0 + \left( \eta' - \eta_0 \right) {\hat k} \right) =  k \, \eta  + \frac{6}{5} k \int_{\eta_{\rm eq}}^\eta d \eta' \, \zeta^L \left(  \left( \eta' - \eta_0 \right) {\hat k} \right).
\end{eqnarray}
Here $\vx_0$ is the location of the detector (that from now on we set to zero for simplicity) and $\hat k$ identifies the direction of motion of the GW. In order to stress that the gravitational potential $\Phi$ has a typical momentum much smaller than $k$, we have used the relation
$\Phi=3\zeta^L/5$. We take $\zeta^L$ Gaussian with the corresponding power spectrum  ${\cal P}^L_\zeta$ scale-invariant.

We have also used the matter-dominated relation, so that $\zeta^L$  is time-independent on linear scales. However, it has a time dependence due to the motion of the GW (that sees a time-varying profile). The solution can be then written as
\begin{eqnarray}
h_\lambda \left( \vec{k} \right) &=&
 \frac{2}{9} \int \frac{d^3 p}{\left( 2 \pi \right)^3} \, \frac{\eta_{\rm eq}}{k^3 \eta^2} \, e_{\lambda,ij}^* \left( {\hat k} \right) {\vec p}_i \vec{p}_j \zeta \left( \vec{p} \right) \zeta \left( \vec{k} - \vec{p} \right) \nonumber\\
& & \times \left[ {\cal I}^* \left( \frac{p}{k}, \frac{\vert\vec{k}-\vec{p}\vert}{k} \right) {\rm e}^{i
k \eta + i \frac{6}{5} k \int_{\eta_{\rm eq}}^\eta  d \eta' \, \zeta^L \left(   \left( \eta' - \eta_0 \right) {\hat k} \right) } +   {\cal I} \left( \frac{p}{k}, \frac{\vert\vec{k}-\vec{p}\vert}{k} \right)  {\rm e}^{-i
k \eta - i \frac{6}{5} k \int_{\eta_{\rm eq}}^\eta  d \eta' \, \zeta^L \left(  - \left( \eta' - \eta_0 \right) {\hat k} \right) }  \right], \nonumber\\
\label{h-zetarealspace}
\end{eqnarray}
where $\zeta$ in the first line is in momentum space, while $\zeta^L$ in the second line is in real space.
One can easily verify that the corresponding Fourier transform is real.

\subsubsection{No effect on  the power spectrum}
In order to see the effect of the propagation onto the power spectrum, it is convenient to define the quantity
\begin{equation}
{\hat Z} \left( \eta,\, \vec{k} \right) \equiv  \frac{6}{5} k \int_{\eta_{\rm eq}}^\eta  d \eta' \, \zeta^L \left(   \left( \eta' - \eta_0 \right) {\hat k} \right).
\end{equation}
Using the results of  section II, we find
\begin{eqnarray}
\left\langle  h_{\lambda_1} \left( \eta ,\, \vec{k} \right)  h_{\lambda_2} \left( \eta ,\, \vec{k}' \right) \right\rangle
 & & =  \frac{\pi^5}{81} \frac{A_s^2}{k_*^2} \,   \frac{\eta_{\rm eq}^2}{k^7 \eta^4} \,  \left( k - 2 k_* \right)^2 \left( k + 2 k_* \right)^2 \,  \theta \left( 2 k_* - k \right)   \, \delta^{(3)} \left( \vec{k} + \vec{k}'  \right)  e_{ab,\lambda_1}^* \left( \vec{k} \right) e_{ab,\lambda_2} \left( \vec{k} \right) \nonumber\\
&&
\times
\Big\langle
\left[ {\cal I}^* \left( \frac{k_*}{k}, \frac{k_*}{k} \right) {\rm e}^{i k \eta + i {\hat Z} \left( \eta,\, \vec{k} \right)  } +
{\cal I} \left( \frac{k_*}{k}, \frac{k_*}{k} \right)
{\rm e}^{-i k \eta - i {\hat Z} \left( \eta,\, -\vec{k} \right)  }
\right]  \nonumber\\
& &
\times \left[
{\cal I}^* \left( \frac{k_*}{k}, \frac{k_*}{k} \right) {\rm e}^{i k \eta + i {\hat Z} \left( \eta,\, - \vec{k} \right)  } +
{\cal I} \left( \frac{k_*}{k}, \frac{k_*}{k} \right) {\rm e}^{-i k \eta - i {\hat Z} \left( \eta,\, \vec{k} \right)  }
\right] \Big\rangle ,
\end{eqnarray}
where we have exploited the fact that the short mode $\zeta$, responsible for the GW production during the radiation-dominated era, and the long mode $\zeta^L$ are not correlated, and therefore
the average splits into the product of two different averages, the one on short-modes being already done.
The contraction of the polarization operators enforces $e_{ab,\lambda_1}^* \left( \vec{k} \right) e_{ab,\lambda_2} \left( \vec{k} \right) = \delta_{\lambda_1,\lambda_2}$. Therefore, we get
\begin{eqnarray}
 P^h \left( k \right)
& & =  \frac{1}{1296} \frac{A_s^2}{k_*^2} \,   \frac{\eta_{\rm eq}^2}{ \eta^4} \,  \left(  \frac{4 k_*^2}{k^2} - 1 \right)^2  \theta \left( 2  - \frac{k}{k_*} \right)   \,   \nonumber\\
&& \times
\Bigg\langle \left[ {\cal I}^* \left( \frac{k_*}{k}, \frac{k_*}{k} \right) {\rm e}^{i \left( k \eta +  {\hat Z} \left( \eta,\vec{k} \right) \right)} +
{\cal I} \left( \frac{k_*}{k}, \frac{k_*}{k} \right)  {\rm e}^{-i \left( k \eta +  {\hat Z} \left( \eta,-\vec{k} \right) \right)}
\right]  \nonumber\\
&& \quad\quad \left[
{\cal I}^* \left( \frac{k_*}{k}, \frac{k_*}{k} \right)  {\rm e}^{i \left( k \eta +  {\hat Z} \left( \eta,-\vec{k} \right) \right)} +
{\cal I} \left( \frac{k_*}{k}, \frac{k_*}{k} \right)  {\rm e}^{-i \left( k \eta +  {\hat Z} \left( \eta,\vec{k} \right) \right)}
\right] \Bigg\rangle .
\end{eqnarray}
In the product, the terms proportional to ${\cal I}^2$ and ${\cal I}^{*2}$ are fast oscillating and average out, while in those proportional to $|{\cal I}|^2$ all phases drop out, and therefore there are no propagation effects in the power spectrum. This leaves us with
\begin{eqnarray}
\overline{{\cal P}^h \left( \eta ,\, k \right)}
& = &  \frac{1}{648} \frac{A_s^2}{k_*^2} \,   \frac{\eta_{\rm eq}^2}{ \eta^4} \,  \left(  \frac{4 k_*^2}{k^2} - 1 \right)^2  \theta \left( 2  - \frac{k}{k_*} \right)  \left[ {\cal I}_c^2 \left( \frac{k_*}{k} ,\, \frac{k_*}{k} \right) +  {\cal I}_s^2  \left( \frac{k_*}{k} ,\, \frac{k_*}{k} \right) \right].
\end{eqnarray}

\subsubsection{The effect of the propagation onto the GW bispectrum}
  \label{BS}
  \noindent
We proceed as in section III, but starting from the solution (\ref{h-zetarealspace}).
From now on we consider the equilateral configuration $\vert \vec{k}_1 \vert = \vert \vec{k}_2 \vert = \vert \vec{k}_3 \vert = k$.
The resulting bispectrum is of the form
\begin{eqnarray}
B_h^{\lambda_i} \left( \eta_i ,\, \vec{k}_i  \right)  & = & {\cal N} \Big<
\left[ {\cal I}^* \left( \frac{k_*}{k}, \frac{k_*}{k} \right)  {\rm e}^{i k \eta_1 + i {\hat Z} \left( \eta_1 ,\, \vec{k}_1 \right) }
+  {\cal I} \left( \frac{k_*}{k}, \frac{k_*}{k} \right)   {\rm e}^{-i k \eta_1 - i {\hat Z} \left( \eta_1 ,\, - \vec{k}_1 \right) }  \right]
\nonumber \\
& & \quad\quad
\left[ {\cal I}^* \left( \frac{k_*}{k}, \frac{k_*}{k} \right) {\rm e}^{i k \eta_2 + i {\hat Z} \left( \eta_2 ,\, \vec{k}_2 \right) }
+  {\cal I}  \left( \frac{k_*}{k}, \frac{k_*}{k} \right){\rm e}^{-i k \eta_2 - i {\hat Z} \left( \eta_2 ,\, - \vec{k}_2 \right) }  \right]
\nonumber\\
& & \quad\quad
\left[ {\cal I}^*\left( \frac{k_*}{k}, \frac{k_*}{k} \right)  {\rm e}^{i k \eta_3 + i {\hat Z} \left( \eta_3 ,\, \vec{k}_3 \right) }
+  {\cal I} \left( \frac{k_*}{k}, \frac{k_*}{k} \right) {\rm e}^{-i k  \eta_3 - i {\hat Z} \left( \eta_3 ,\, - \vec{k}_3 \right) }  \right]\Big>,
\end{eqnarray}
where we have defined
\begin{equation}
{\cal N} \equiv  \frac{ \theta \left( \sqrt{3} k_* - k \right) }{k^6}
\times \frac{A_s^3 \, \eta_{\rm eq}^3}{k_*^3 \, \eta_1^2 \eta_2^2 \eta_3^2} \times \frac{1024 \pi^3 }{729}
\times  {\cal D}_{\lambda_i}  \left(  \sqrt{\frac{3 k_*^2}{k^2}-1} \right)^{-1/2} \times \frac{1}{8}.
\label{calN}
\end{equation}
 We also set all times equal, since small relative variations of the three times (hours vs. the age of the universe) will not affect our result.
By defining
\begin{equation}
{\cal G}_{c_1,c_2,c_3} \left[ \vec{k}_1 ,\, \vec{k}_2 ,\, \vec{k}_3 \right] \equiv \left\langle
{\rm e}^{i c_1 {\hat Z} \left( \eta ,\, \vec{k}_1 \right)}
{\rm e}^{i c_2 {\hat Z} \left( \eta ,\, \vec{k}_2 \right)} {\rm e}^{i c_3 {\hat Z} \left( \eta ,\, \vec{k}_3 \right)}
\right\rangle
\end{equation}
we can express our result as
\begin{eqnarray}
\label{pp}
B_h^{\lambda_i} \left( \eta ,\, \vec{k}_i  \right)
&=& {\cal N} \Bigg\{ {\cal I}^{* 3} \, {\rm e}^{3 i k \eta} \,
{\cal G}_{+++} \left[ \vec{k}_1 ,\, \vec{k}_2 ,\, \vec{k}_3 \right]  \nonumber\\
& & \quad + {\cal I}^{* 2} \, {\cal I} \, {\rm e}^{i k \eta} \left[
{\cal G}_{++-} \left[ \vec{k}_1 ,\, \vec{k}_2 ,\, -\vec{k}_3 \right]  +
{\cal G}_{+-+} \left[ \vec{k}_1 ,\, -\vec{k}_2 ,\, \vec{k}_3 \right]  +
{\cal G}_{-++} \left[ -\vec{k}_1 ,\, \vec{k}_2 ,\, \vec{k}_3 \right]  \right] \nonumber\\
& & \quad + {\cal I}^* \, {\cal I}^2 \, {\rm e}^{-i k \eta} \left[
{\cal G}_{+--} \left[ \vec{k}_1 ,\, -\vec{k}_2 ,\, -\vec{k}_3 \right]  +
{\cal G}_{-+-} \left[ -\vec{k}_1 ,\, \vec{k}_2 ,\, -\vec{k}_3 \right]  +
{\cal G}_{--+} \left[ -\vec{k}_1 ,\, -\vec{k}_2 ,\, \vec{k}_3 \right]  \right] \nonumber\\
& & \quad +   {\cal I}^3 \, {\rm e}^{-3 i k \eta} \,
{\cal G}_{---} \left[ -\vec{k}_1 ,\, -\vec{k}_2 ,\, -\vec{k}_3 \right] \Bigg\},
\end{eqnarray}
where, having all the same argument,  we set ${\cal I} \left( \frac{k_*}{k} ,\, \frac{k_*}{k} \right) \rightarrow {\cal I}$ .
Next, we use the identity
\begin{eqnarray}
\left\langle {\rm e}^{\varphi_1} {\rm e}^{\varphi_2} {\rm e}^{\varphi_3} \right\rangle
 = {\rm e}^{\frac{\langle\varphi_1^2\rangle}{2} + \frac{\langle\varphi_2^2\rangle}{2} +  \frac{\langle\varphi_3^2\rangle}{2} +
\langle\varphi_1 \varphi_2\rangle + \langle\varphi_1 \varphi_3\rangle + \langle\varphi_2 \varphi_3\rangle }
\end{eqnarray}
valid for a general Gaussian operator, and write
\begin{equation}
{\cal G}_{c_1,c_2,c_3} \left[ \vec{k}_1 ,\, \vec{k}_2 ,\, \vec{k}_3 \right]
= {\rm exp}^{
-\frac{1}{2}c_1^2 {\cal C} \left( \vec{k}_1 ,\, \vec{k}_1 \right)
-\frac{1}{2}c_2^2 {\cal C} \left( \vec{k}_2 ,\, \vec{k}_2 \right)
-\frac{1}{2} c_3^2 {\cal C} \left( \vec{k}_3 ,\, \vec{k}_3 \right)
- c_1 c_2 {\cal C} \left( \vec{k}_1 ,\, \vec{k}_2 \right)
- c_1 c_3 {\cal C} \left( \vec{k}_1 ,\, \vec{k}_3 \right)
- c_2 c_3 {\cal C} \left( \vec{k}_2 ,\, \vec{k}_3 \right)
}
\end{equation}
where we defined
\begin{equation}
{\cal C} \left( \vec{k}_1 ,\, \vec{k}_2 \right) \equiv
\left\langle {\hat Z} \left( \eta ,\, \vec{k}_1 \right) \,  {\hat Z} \left( \eta ,\, \vec{k}_2 \right) \right\rangle=\frac{36}{25} k_1 k_2
\int_{\eta_{\rm eq}}^\eta  d \eta' \,  \int_{\eta_{\rm eq}}^\eta  d \eta'' \,
\left\langle \zeta^L \left(   \left( \eta' - \eta_0 \right) {\hat k}_1 \right)
\zeta^L \left(   \left( \eta''- \eta_0 \right) {\hat k}_2 \right) \right\rangle .
\end{equation}
We Fourier transform
\begin{equation}
\zeta^L \left(  \left( \eta - \eta_0 \right) {\hat k} \right) =
\int \frac{d^3 p}{\left( 2 \pi \right)^3 } \, {\rm e}^{ i \left( \eta - \eta_0 \right) {\hat k} \cdot \vec{p} } \zeta^L \left( \vec{p} \right)
\end{equation}
and compute the correlator
\begin{eqnarray}
\left\langle
\zeta^L \left(  \left( \eta' - \eta_0 \right) {\hat k}_1 \right)
\zeta^L \left(  \left( \eta'' - \eta_0 \right) {\hat k}_2 \right)
\right\rangle &=&
\int \frac{d^3 p_1 d^3 p_2}{\left( 2 \pi \right)^6} \,
{\rm e}^{i \left( \eta' - \eta_0 \right) {\hat k}_1 \cdot \vec{p}_1}
{\rm e}^{i \left( \eta'' - \eta_0 \right) {\hat k}_2 \cdot \vec{p}_2}
\left\langle \zeta^L \left( \vec{p}_1 \right)  \zeta^L \left( \vec{p}_2 \right) \right\rangle \nonumber\\
 && \!\!\!\!\!\!\!\!  \!\!\!\!\!\!\!\!  \!\!\!\!\!\!\!\!  \!\!\!\!\!\!\!\!
 =  \int \frac{d^3 p_1 d^3 p_2}{\left( 2 \pi \right)^6} \,
{\rm e}^{i \left( \eta' - \eta_0 \right) {\hat k}_1 \cdot \vec{p}_1}
{\rm e}^{i \left( \eta'' - \eta_0 \right) {\hat k}_2 \cdot \vec{p}_2}
\left( 2 \pi \right)^3 \delta^{(3)} \left( \vec{p}_1 + \vec{p}_2 \right) \frac{2 \pi^2}{p_1^3} \, {\cal P}_\zeta^L \left( p_1 \right) \nonumber\\
 && \!\!\!\!\!\!\!\!  \!\!\!\!\!\!\!\!  \!\!\!\!\!\!\!\!  \!\!\!\!\!\!\!\!
 =  \int \frac{d^3 p}{4 \pi p^3} \,
{\rm e}^{i \left( \eta' - \eta_0 \right) {\hat k}_1 \cdot \vec{p}}
{\rm e}^{-i \left( \eta'' - \eta_0 \right) {\hat k}_2 \cdot \vec{p}}  \, {\cal P}_\zeta^L \left( p \right).
\end{eqnarray}
Therefore
\begin{equation}
{\cal C} \left( \vec{k}_1 ,\, \vec{k}_2 \right) = \frac{36}{25} k_1 k_2
\int_{\eta_{\rm eq}}^\eta  d \eta' \,  \int_{\eta_{\rm eq}}^\eta  d \eta'' \,
\int \frac{d^3 p}{4 \pi p^3} \,
{\rm e}^{i \left( \eta' - \eta_0 \right) {\hat k}_1 \cdot \vec{p}}
{\rm e}^{-i \left( \eta'' - \eta_0 \right) {\hat k}_2 \cdot \vec{p}}  \, {\cal P}_\zeta^L \left( p \right)
\end{equation}
or, changing the order of times in the exponent (so to have a positive time difference in the bracket)
\begin{equation}
{\cal C} \left( \vec{k}_1 ,\, \vec{k}_2 \right) = \frac{36}{25} k_1 k_2
\int_{\eta_{\rm eq}}^\eta  d \eta' \,  \int_{\eta_{\rm eq}}^\eta  d \eta'' \,
\int \frac{d^3 p}{4 \pi p^3} \,
{\rm e}^{-i \left( \eta_0 - \eta' \right) {\hat k}_1 \cdot \vec{p}}
{\rm e}^{i \left( \eta_0 - \eta'' \right) {\hat k}_2 \cdot \vec{p}}  \, {\cal P}_\zeta^L \left( p \right).
\end{equation}
Next, we use the expansion in spherical harmonics
\begin{eqnarray}
{\rm e}^{i \vec{k} \cdot \vec{r}}
&=& 4 \pi \sum_{\ell=0}^\infty \sum_{m=-\ell}^{\ell} \, i^\ell \, j_\ell \left( k \, r \right) Y_{\ell m} \left( {\hat r} \right)  Y_{\ell m}^* \left( {\hat k} \right)
\end{eqnarray}
so that
\begin{eqnarray}
{\cal C} \left( \vec{k}_1 ,\, \vec{k}_2 \right) &=& \frac{36}{25} k_1 k_2 \int_{\eta_{\rm eq}}^\eta  d \eta' \,  \int_{\eta_{\rm eq}}^\eta  d \eta'' \;
 \left( 4 \pi \right)^2
\sum_{\ell=0}^\infty \sum_{m=-\ell}^{\ell} \sum_{\ell'=0}^\infty \sum_{m'=-\ell'}^{\ell'} \left( i \right)^{-\ell + \ell'}
  \,  Y_{\ell m}^* \left( {\hat k}_1 \right) \,  Y_{\ell' m'} \left( {\hat k}_2 \right)  \nonumber\\
& & \times  \int \frac{d p}{p}  \, {\cal P}_\zeta^L \left( p \right)
  \,  j_\ell \left(  \left( \eta_0 - \eta' \right) p \right) \,  j_{\ell'} \left(  \left( \eta_0 - \eta'' \right) p \right)
   \nonumber\\
& & \times \int \frac{d \Omega_{\hat p}}{4 \pi} \,
Y_{\ell m} \left( {\hat p} \right) \,  Y_{\ell' m'}^* \left( {\hat p} \right).
\end{eqnarray}
Being the last line equal to $\delta_{\ell \ell'} \delta_{m m'}/4 \pi$, we obtain
\begin{eqnarray}
{\cal C} \left( \vec{k}_1 ,\, \vec{k}_2 \right) &=& \frac{36}{25} k_1 k_2 \int_{\eta_{\rm eq}}^\eta  d \eta' \,  \int_{\eta_{\rm eq}}^\eta  d \eta'' \;
 4 \pi
\sum_{\ell=0}^\infty \sum_{m=-\ell}^{\ell}
  \,  Y_{\ell m}^* \left( {\hat k}_1 \right) \,  Y_{\ell m} \left( {\hat k}_2 \right)  \nonumber\\
& & \times  \int \frac{d p}{p}  \, {\cal P}_\zeta^L \left( p \right)
  \,  j_\ell \left(  \left( \eta_0 - \eta' \right) p \right) \,  j_{\ell} \left(  \left( \eta_0 - \eta'' \right) p \right).
\end{eqnarray}
Notice that the integral diverges logarithmically due to the $\ell =0$ term, that is due to the constant zero mode of the gravitational potential. Being it unphysical, we follow the procedure of Ref. \cite{hu} and remove the monopole from the sum. For the other multiples, we use the approximation (which, strictly speaking, requires $\ell \gg 1$) \cite{vg}
\begin{equation}
\int_0^\infty d p \, p^2 \, f \left( p \right) j_\ell \left( p \eta \right) j_{\ell} \left( p \eta' \right) \simeq
\frac{\pi}{2 \eta^2}  f \left( \frac{\ell + 1/2}{\eta} \right) \, \delta_D \left( \eta - \eta' \right)
\end{equation}
where for us
\be
f(p)=\frac{{\cal P}_\zeta^L(p)}{p^3}=\frac{A_L}{p^3}.
\ee
With this in mind, we can write
\begin{eqnarray}
{\cal C} \left( \vec{k}_1 ,\, \vec{k}_2 \right) &=& \frac{36}{25} k_1 k_2 \int_{\eta_{\rm eq}}^\eta  d \eta' \,  \int_{\eta_{\rm eq}}^\eta  d \eta'' \;
 4 \pi
\sum_{\ell=1}^\infty \sum_{m=-\ell}^{\ell}
  \,  Y_{\ell m}^* \left( {\hat k}_1 \right) \,  Y_{\ell m} \left( {\hat k}_2 \right)  \nonumber\\
& & \times \frac{\pi}{2} \frac{1}{\left( \eta_0 - \eta' \right)^2}
\frac{A_L \left( \eta_0 - \eta' \right)^3 }{\left( \ell + \frac{1}{2} \right)^3} \, \delta_D \left( \eta' - \eta'' \right)
\end{eqnarray}
which gives
\begin{eqnarray}
{\cal C} \left( \vec{k}_1 ,\, \vec{k}_2 \right) &=& \frac{36}{25} k_1 k_2 \times 2 \pi^2 \, \int_{\eta_{\rm eq}}^\eta  d \eta' \,  \left( \eta_0 - \eta' \right)
\sum_{\ell=1}^\infty \sum_{m=-\ell}^{\ell}  \frac{A_L  }{\left( \ell + \frac{1}{2} \right)^3}
  \,  Y_{\ell m}^* \left( {\hat k}_1 \right) \,  Y_{\ell m} \left( {\hat k}_2 \right).
\end{eqnarray}
In the time integration we now put $\eta_0$ also as extremum of integration and disregard the equality time. Furthermore, in the spherical harmonics we can always orient ${\hat k}_1$ along the $z-$axis, so that the sum involves only the $m=0$ terms. We then note that $\cos {\hat k}_2$ becomes ${\hat k}_1 \cdot {\hat k}_2 \equiv \mu$. With this convention, then
\begin{equation}
Y_{\ell m}^* \left( {\hat k}_1 \right) \,  Y_{\ell m} \left( {\hat k}_2 \right) = \delta_{m0} \frac{2 \ell + 1}{4 \pi}
P_\ell \left( 1 \right) P_\ell \left( \mu \right) =  \delta_{m0} \frac{2 \ell + 1}{4 \pi}
P_\ell \left( \mu \right) ,
\end{equation}
where in the last step we used $P_\ell \left( 1 \right) =1$.
All this gives
\begin{eqnarray}
{\cal C} \left( \vec{k}_1 ,\, \vec{k}_2 \right) = A_L \, k_1 \, k_2 \, \eta_0^2 \times
 \frac{18 \, \pi}{25} \: \sum_{\ell=1}^\infty   \frac{ P_\ell \left( {\hat k}_1 \cdot {\hat k}_2 \right) }{\left( \ell + \frac{1}{2} \right)^2} \equiv  A_L \, k_1 \, k_2 \, \eta_0^2 \times {\cal S} \left(  {\hat k}_1 \cdot {\hat k}_2 \right).
\end{eqnarray}
The equilateral bispectrum involves the following momentum configurations
\begin{eqnarray}
i = j \;:\;\;\; {\cal C} \left( \vec{k}_i ,\, \vec{k}_j \right) &=&  A_L \, k^2 \, \eta_0^2 \times  {\cal S} \left(  1 \right) \simeq \frac{18 \, \pi}{25} \left( \frac{\pi^2}{2} - 4 \right)  \; A_L \, \left( k \, \eta_0 \right)^2
\simeq 2.11 \; A_L \, \left( k \, \eta_0 \right)^2,  \nonumber\\
i \neq j \;:\;\;\; {\cal C} \left( \pm \vec{k}_i ,\, \pm \vec{k}_j \right) &=&  A_L \, k^2 \, \eta_0^2 \times  {\cal S} \left(  - 1 / 2 \right) \simeq -0.50  \; A_L \, \left( k \, \eta_0 \right)^2, \nonumber\\
i \neq j \;:\;\;\; {\cal C} \left( \pm \vec{k}_i ,\, \mp \vec{k}_j \right) &=&  A_L \, k^2 \, \eta_0^2 \times  {\cal S} \left(   1 / 2 \right) \simeq 0.37  \; A_L \, \left( k \, \eta_0 \right)^2.
\label{calC-res}
\end{eqnarray}
Inserting them into Eq. (\ref{pp}), then we obtain
\begin{eqnarray}
B_h^{\lambda_i} \left( \eta_0 ,\, \vec{k}_i  \right)
&=& {\cal N} \Bigg\{ {\cal I}^{* 3} \, {\rm e}^{3 i k \eta_0} \,
{\rm e}^{-1.67 \,  A_L \, \left( k \, \eta_0 \right)^2 }
 + {\cal I}^{* 2} \, {\cal I} \, {\rm e}^{i k \eta_0}
\left[ {\rm e}^{-1.93 \,  A_L \, \left( k \, \eta_0 \right)^2 } \times 3  \right] \nonumber\\
& & \quad + {\cal I}^* \, {\cal I}^2 \, {\rm e}^{-i k \eta_0}
\left[ {\rm e}^{-1.93 \,  A_L \, \left( k \, \eta_0 \right)^2 } \times 3  \right] +   {\cal I}^3 \, {\rm e}^{-3 i k \eta_0} \,
{\rm e}^{-1.67 \,  A_L \, \left( k \, \eta_0 \right)^2 }  \Bigg\}.
\end{eqnarray}
The first and last term are the least suppressed ones, giving
\begin{eqnarray}
B_h^{\lambda_i} \left( \eta_0 ,\, \vec{k}_i  \right)
&=& {\cal N} \, {\rm e}^{-1.67 \,  A_L \, \left( k \, \eta_0 \right)^2 } \,
\left[ {\cal I}^{* 3} \, {\rm e}^{3 i k \eta_0}  +   {\cal I}^3 \, {\rm e}^{-3 i k \eta_0} \,  \right].
\end{eqnarray}
Therefore we have
\begin{eqnarray}
\frac{B_h^{\lambda_i} \left( \eta_0 ,\, \vec{k}_i  \right) \Big\vert_{\rm inhom.} }
{B_h^{\lambda_i} \left( \eta_0 ,\, \vec{k}_i  \right) \Big\vert_{\rm no \; inhom.} }
&=& {\rm e}^{-1.67 \,  A_L \, \left( k \, \eta_0 \right)^2 } .\,
\end{eqnarray}
We further define and compute the rms of the (relative) time delay $d$, see \cite{hu}, as
\begin{eqnarray}
d_{\rm rms}^2 &=& \frac{4}{\eta_0^2} \left\langle \int d \eta' d \eta'' \Phi \left( \eta' \right)  \Phi \left( \eta'' \right) \right\rangle = \frac{4}{\eta_0^2 k^2} \times \frac{9}{25} k^2  \left\langle \int d \eta' d \eta'' \zeta_L \left( \eta' \right)  \zeta_L \left( \eta'' \right) \right\rangle \nonumber\\
&=&  \frac{1}{\eta_0^2 \, k^2} \, {\cal C} \left( \vec{k} ,\, \vec{k} \right) =  2.11 \, A_L,
\end{eqnarray}
where in the last step we have used the first of Eqs.~(\ref{calC-res}). With this, the above ratio finally rewritten as
\begin{eqnarray}
\frac{B_h^{\lambda_i} \left( \eta_0 ,\, \vec{k}_i  \right) \Big\vert_{\rm inhom.} }
{B_h^{\lambda_i} \left( \eta_0 ,\, \vec{k}_i  \right) \Big\vert_{\rm no \; inhom.} }
&=& {\rm e}^{-1.67 \,  A_L \, \left( k \, \eta_0 \right)^2  \, \frac{d_{\rm rms}^2}{2.11 \, A_L} } =
 {\rm e}^{-0.8  \,  k^2 \eta_0^2  \, d_{\rm rms}^2 }.
\end{eqnarray}
The conclusion of this calculation shows that the primordial bispectrum is not preserved after the propagation of the GWs and is largely suppressed being $k \eta_0  \, d_{\rm rms}\sim 10^9$, if we take $k\sim 10^{-3}$ Hz.

It is interesting to notice that in the squeezed limit $k_1\sim k_2\gg k_3$, the bispectrum should reduce to the average of the short-mode power spectrum in a background modulated by the long mode $k_3$. Indeed, repeating the steps we have performed above for the squeezed limit, it is easy to see that the bispectrum reduces to
\begin{eqnarray}
B_h^{\lambda_i} \left( \eta ,\, \vec{k}_i  \right) \propto
  |{\cal I}_1|^2 \left( {\cal I}_3^* {\rm e}^{i \eta k_3} +  {\cal I}_3 {\rm e}^{-i \eta k_3}  \right)
{\rm e}^{-\frac{1}{2}  k_3^2 \, \eta_0^2 \, d_{\rm rms}^2 },
\end{eqnarray}
where we have kept the term which is least suppressed by the propagation.  This indeed is proportional to the power spectrum of the short mode times a suppression from the long mode due to the fact that the average over the long mode has to be performed over many directions. Unfortunately, the suppression is still sizeable. We conclude that propagation effects are present for arbitrary shapes.

\section{Conclusions}
\noindent
In this paper we have investigated the capability of the LISA project to detect the non-Gaussian GWs generated during the physical process giving rise to
PBHs. The latter are formed through the collapse of the initial large scalar perturbations generated during inflation at the moment when they re-enter the horizon in the radiation-dominated phase. The very same scalar perturbations act as a second-order source for the GWs, which therefore are born non-Gaussian.

If the corresponding PBH masses are in the range around $\sim 10^{-12} M_\odot$, not only these PBHs may form the totality of the dark matter, but also the present frequency of the corresponding GWs happens to be in the ballpark of the mHz, precisely
where LISA has its maximum sensitivity. We have shown that LISA will detect such signal, but  only  through the two-point correlator. For the first time we have proven that the relatively short observation time (as compared with the age of the universe) and the propagation effects of the GWs in the perturbed universe make the three-point correlator not measurable in its simplest form. The propagation effects have been obtained  by solving the equation for the GWs in the geometrical optic limit, making explicit the suppression factor arising when averaging over the directions of propagation. We should stress, however, that the power spectrum in this regime is unaffected and we will devote further investigations to analyse the implications of the propagation effects going beyond the geometrical optic limit.

Finally, the absence of a GW signal will tell us that PBHs of masses around $\sim 10^{-12} M_\odot$ are not the dark matter we observe in the universe. As already mentioned in the introduction,  however, a detection of a GW signal will be still compatible with our universe populated by such PBHs in smaller abundances.
Of course, our  considerations hold within the set-up considered in this paper, leaving  open the possibility that the PBHs originate from mechanisms.

\bigskip
\noindent
\centerline{\bf Note added}
\vskip 0.5cm
\noindent
While completing this work, Ref.~\cite{sasakinew} appeared which also noted that PBHs as dark matter with masses around $10^{-12} M_\odot$ are associated to GWs which can be detected by LISA. However, there the investigation is limited to the GWs power spectrum (with non-Gaussianity in the scalar perturbations), and the study of the bispectrum and its detectability is not present. Ref.~\cite{sasakinew} considered the case in which the scalar perturbations have a non-Gaussian statistics of the local type. This non-Gaussianity results in a greater PBH production (at equal power spectrum of the scalar perturbations). This therefore results in stricter bounds on the scalar power spectrum, and on a smaller GW production, which is however still detectable at LISA~\cite{sasakinew}. The GW production from non-Gaussian scalar perturbations leading to PBH was also previously studied in Ref.~\cite{compagno}, that assumed a non-Gaussian parameterisation of the  probability density function of the scalar perturbations, and in Ref.~\cite{gbp}, that studied the case of a $\chi^2$ distribution, which emerges in models of axion inflation with the inflaton coupled to gauge fields. 
\bigskip
\vskip 0.4cm
\noindent
\centerline{\bf  Acknowledgments}
\vskip 0.5cm
\noindent
We warmly thank the anonymous referee and A. Lewis for asking about the propagation effect, which has led to the revised version.  We also thank A. Lewis for inputs about the non-measurability of the bispectrum.    N.B. acknowledges partial financial support by the ASI/INAF Agreement I/072/09/0 for the Planck LFI Activity of Phase E2. He also  acknowledges financial support   by  ASI Grant 2016-24-H.0.
A.R. is  supported by the Swiss National Science Foundation (SNSF), project {\sl The Non-Gaussian Universe and Cosmological Symmetries}, project number: 200020-178787. %
This research was supported in part by Perimeter Institute for Theoretical Physics. Research at
Perimeter Institute is supported by the Government of Canada through the Department of
Innovation, Science and Economic Development and by the Province of Ontario through the
Ministry of Research, Innovation and Science.


\appendix

\section{Absence of constraints on PBHs as Dark Matter for $10^{-13}-10^{-11}\,M_\odot$}
\label{app:PBH constraints}
\noindent
In this Appendix we review the reasons why the PBH mass region of interest to our paper, between $10^{-13}$ and $10^{-11}\,M_\odot$, is currently unconstrained by observational constraints\footnote{We thank A. Katz for illuminating discussions on the microlensing and neutron star constraints on the PBH abundance.}.

One controversial bound falling in this mass window is the dynamical constraint related to the effect of PBHs on neutron stars in their surroundings.
For low masses $\MPBH\sim 10^{-14}-10^{-13}\,M_\odot$, PBHs could be dense enough to collide with White Dwarves (WD), compact stars which could be dynamically heated by the passage PBHs and explode as supernovae. The observed WD density prevents PBHs in this mass range from forming all of DM \cite{Graham:2015apa}.\\
Neutron Stars (NS) are much smaller and denser than WD, and a PBH could be captured by a NS within a longer time through multiple oscillations around it.
This process is eased if the velocity of the PBHs is small.
The authors of Ref.~\cite{neut} consider NS in the cores of globular clusters, where the velocity dispersion is of order $\mathcal O(10)$ km/s, and derive a constraint in the range $\MPBH \sim 10^{-14}-10^{-10}\,M_\odot$.
The critical assumption that they make concerns the DM density in globular clusters, which is needed to be $10^2$ to $10^4$ times larger than the average $0.3$ GeV/cm$^3$ in the halo in order to yield a constraint.
For more conservative assumptions this bound disappears.

Another bound that we cut at $\MPBH\sim 10^{-11}\,M_\odot$ comes from the observations with the Hyper Suprime-Cam (HSC) of the Subaru telescope of stars in the Andromeda galaxy M31, at 770 kpc from us.
Ref.~\cite{hsc} analysed its data in the search for microlensing events of the stars measured by HSC in the optical window, and derived a constraint in the range $10^{-13}-10^{-6} \,M_\odot$.\\
A first major effect which was unaccounted for in the analysis is the finite size of the sources.
The projected size of the main sequence star onto the lensing plane radius is larger by more than order of magnitude than the Einstein radius for PBHs of mass $10^{-11}\,M_\odot$.
Therefore the magnification of the star drops well below 10\%, a factor of 10 below what  is needed to detect a signal.\\
Another important phenomenon which further weakens the microlensing constraints below $\sim 10^{-10}\,M_\odot$ and makes them disappear below $\sim 10^{-11}$ is the so called wave effect \cite{Gould:1991td,Nakamura:1997sw,hsc1,k}.
When the observed light has a wavelength $\lambda$ larger  than the Schwarzschild radius $\rS$ of the lens, then the framework of geometrical optics does not provide a good description of the system, and the diffraction effects give small or vanishing magnification of the source.
It is interesting to understand why $\rS$ is a relevant quantity in this discussion, since the deflected images of the source travel at a larger distance $\rE\gg \rS$ from the lens.
The explanation is related to the expression of the Einstein radius $\rE\sim \sqrt{\rS D}$.
We denote by $D$ the distance between the observer and the lens, which we assume for simplicity to be of the same order as the distance between the lens and the source.
One can interpret the travel of the lensed rays at opposite sides of the lens as a double slit interference experiment.
The first maxima of the diffracted pattern occur at angles $\theta_1\sim \lambda /\rE$, where $\rE$ is the equivalent of the distance between the slits.
With respect to the line of sight to the PBH, the deflected rays are at an angle $\theta_{{\text{\tiny S}}} \sim \rE/D$.
If $\theta{{\text{\tiny S}}} \gg \theta_1$, then the interference pattern is not resolvable, the wave effects of light diffusion are negligible and the PBH appreciably magnifies the background objects.
On the other hand, if $\theta_{{\text{\tiny S}}}<\theta_1$, then the interference pattern is visible, and the geometrical optics approximation breaks down.
The condition for magnification $\theta_{{\text{\tiny S}}} \gtrsim \theta_1$ yields $\lambda \lesssim \rE^2/D\sim \rS$.

Because of these reasons, we cut the constraints from the Subaru HSC observations below $\MPBH\sim 10^{-11}\,M_\odot$.

\section{Computation of ${\cal D}_{\lambda_i}$}
 \label{app:calD}
\renewcommand\theequation{\Alph{section}.\arabic{equation}}

\noindent
In this Appendix we present the results for the contractions
\begin{equation}
 {\cal D}_{\lambda_i} \left( {\hat k}_i ,\, r_i \right)  \equiv e_{ab,\lambda_1}^* \left( {\hat k}_1 \right) e_{cd,\lambda_2}^* \left( {\hat k}_2 \right) e_{ef,\lambda_3}^* \left( {\hat k}_3 \right)
\left\{ \left[ \vec{q}_a \, \vec{q}_b  \left( \vec{q} - {\hat k}_1 \right)_c   \left( \vec{q} - {\hat k}_1 \right)_d  \vec{q}_e \, \vec{q}_f   \right]_I + \left[ {\rm same} \right]_{II} \right\},
\label{calD}
\end{equation}
where the momenta $\vec{k}_i$ are given in Eq.(\ref{3frame}), while the momenta $\vec{q}_{I,II}$ are given in Eq.(\ref{p1-2}). We construct the polarization operators as it is standard, following the notation of Ref.~\cite{snr}. We define the two unit vectors orthogonal to the GW momentum through an external fixed unit vector ${\hat e}_z$, that we choose to be the unit vector along the third axis
\begin{equation}
{\hat u} \equiv \frac{{\hat k} \cdot {\hat e}_z}{\left\vert {\hat k} \cdot {\hat e}_z \right\vert} ,\;\;
{\hat v} \equiv {\hat k} \cdot {\hat u},
\end{equation}
(clearly, any other fixed vector can equivalently be chosen.)  Starting from this choice, we then introduce the left handed and right handed polarization operators
\begin{equation}
e_{ab,{\rm R}} \equiv \frac{{\hat u}_a + i {\hat v}_a}{\sqrt{2}}  \frac{{\hat u}_b + i {\hat v}_b}{\sqrt{2}},\;\;
e_{ab,{\rm L}} \equiv \frac{{\hat u}_a - i {\hat v}_a}{\sqrt{2}}  \frac{{\hat u}_b - i {\hat v}_b}{\sqrt{2}}.
\end{equation}
These operators are symmetric, transverse, traceless, and normalized according to $e_{ab,\lambda}^* e_{ab,\lambda} = 1$. They are also related to the operators $e_+,\, e_\times$ as
\begin{equation}
e_{ab,{\rm R/L}} = \frac{e_{ab,+} \pm i \, e_{ab,\times}}{\sqrt{2}}.
\end{equation}
The contractions   can be readily obtained from these expressions for the polarization operators, and from Eqs.(\ref{p1-2}). The resulting expression are rather lengthy. We report them here in the isosceles case $r_1 = r_2$, where these expressions acquire a more compact form
\begin{eqnarray}
{\cal D}_{\rm RRR} = {\cal D}_{\rm LLL} &=& \frac{1}{256} \Bigg[ \frac{32 r_1^3}{(2 r_1+r_3)^3}-\frac{24 \left(3
r_1^2+8\right)}{(2 r_1+r_3)^2}+\frac{32
\left(r_1^2+4\right) r_3}{r_1^5}+\frac{32
\left(r_1^2-1\right)^2}{r_1^3 (2
r_1-r_3)} \nonumber\\
& & \quad\quad +\frac{32 \left(2 \left(r_1^2+6\right)
r_1^2+9\right)}{r_1^3 (2
r_1+r_3)}-\frac{\left(r_1^4+24 r_1^2+16\right)
r_3^2}{r_1^6}-\frac{4 \left(33 r_1^4+24
r_1^2+16\right)}{r_1^6}-32 \Bigg], \nonumber\\
{\cal D}_{\rm LRR} = {\cal D}_{\rm RLL} &=&  {\cal D}_{RLR} = {\cal D}_{LRL} =   \frac{\left(r_1^2-4\right)^2 \left(8 r_1^4-4 r_1^2
   \left(r_3^2+4\right)+r_3^4+4 r_3^2\right)}{256
   r_1^6 \left(4 r_1^2-r_3^2\right)} , \nonumber\\
{\cal D}_{\rm RRL} = {\cal D}_{\rm LLR} &=& \frac{1}{256} \Bigg[ -\frac{32 r_1^3}{(r_3-2 r_1)^3}-\frac{24 \left(3
   r_1^2+8\right)}{(r_3-2 r_1)^2}-\frac{32
   \left(r_1^2+4\right) r_3}{r_1^5}-\frac{32 \left(2
   \left(r_1^2+6\right) r_1^2+9\right)}{r_1^3 (r_3-2
   r_1)} \nonumber\\
& & \quad\quad
+\frac{32 \left(r_1^2-1\right)^2}{r_1^3 (2
   r_1+r_3)}-\frac{\left(r_1^4+24 r_1^2+16\right)
   r_3^2}{r_1^6}-\frac{4 \left(33 r_1^4+24
   r_1^2+16\right)}{r_1^6}-32 \Bigg] .
\end{eqnarray}
We note that the contractions are invariant under parity (L $ \leftrightarrow $ R). Moreover, we note that
LRR = RLR (due to the fact that $r_2 = r_3$).

\section{Redshift} \label{app:redshift}
\noindent
The bispectrum (\ref{bispectrum}) can be rewritten as
\begin{equation}
\big< h(\eta,\vk_1) h(\eta,\vk_2) h(\eta,\vk_3) \big>' = \frac{{\hat B}}{\eta^3} ,
\end{equation}
where we have factored out the conformal time $\eta$, that accounts for the decrease of the amplitude of the GW mode functions while inside the horizon during radiation domination. Since, in general, the GW amplitude scales as the inverse power of the scale factor inside the horizon, $\med{hhh} \sim 1/a^3$,  the bispectrum evaluated today at $\eta_0$ is given by:
\begin{equation}
	\label{Eq.red}
	\big< h(\eta_0,\vk_1) h(\eta_0,\vk_2) h(\eta_0,\vk_3) \big>'  = \bigg(\frac{a_\text{\tiny eq}}{a_0}\bigg)^3 \big< h(\eta_\text{\tiny eq},\vk_1) h(\eta_\text{\tiny eq},\vk_2) h(\eta_\text{\tiny eq},\vk_3) \big>' = \bigg(\frac{a_\text{\tiny eq}}{a_0}\bigg)^3 \frac{1}{\eta_\text{\tiny eq}^3} {\hat B} ,
\end{equation}
where $\eta_\text{\tiny eq}$ represents the conformal time at the radiation-matter equality.
During the radiation-dominated  phase the scale factor goes as $a \sim \eta$, and so we can write:
\begin{equation}
	a_\text{\tiny eq} = a_{k} \cdot \frac{\eta_\text{\tiny eq}}{\eta_k} ,
\end{equation}
where the subscript ${}_k$ denotes the moment at which the comoving momentum $k$ re-enters the horizon. The crossing happens when $k = aH$
such that during the radiation era   we have $\eta_k = 1/k = 1/2\pi f$.
Appendix G of Ref.~\cite{Lisa} gives
\begin{equation}
	\frac{a_{\rm 0}}{a_k} \sim 2 \cdot 10^{17} \frac{f}{10^{-3} \rm Hz},
\end{equation}
that can be used to derive the present bispectrum
\begin{equation}
	\label{redshift}
	\big< h(\eta_0,\vk_1) h(\eta_0,\vk_2) h(\eta_0,\vk_3) \big>'  = (\pi \cdot 10^{-20} {\rm Hz})^3 \cdot {\hat B} (\vk_1,\vk_2,\vk_3).
\end{equation}
Although in this discussion, for brevity, the equal time bispectrum had been considered, this relation applies also to the unequal time bispectrum considered in the main text.
\noindent

\section{LISA response functions for the bispectrum} \label{app:LISA}
\noindent
Even though we have shown that propagation effects make the primordial bispectrum of GWs highly suppressed, it is nevertheless interesting to analyse the LISA response functions for the three-point correlator. Implicit expressions for the  response functions are given in Ref. \cite{snr}; here we improved over those results by providing explicit simple approximations of the response functions close to their maxima.

The LISA constellation consists of three satellites placed at the vertices of an equilateral triangle of side $L = 2.5 \cdot 10^6 \, {\rm km}$ (the actual distance slightly varies during the orbit; this effect is disregarded in our computations). Laser light is sent from each satellite to the other two, so that each vertex acts as a time delay interferometer. The three measurements are not noise-orthogonal, as any two interferometers share one arm. The noise covariance matrix can however be diagonalized to provide noise-orthogonal combinations. We consider the three linear combinations A,E,T introduced in Ref. \cite{corn}. If we think at LISA as three interferometers, one centered at each satellite, and we denote by $\sigma_{X,Y,Z}$ the time delay measured by each satellite, the A,E,T channel are the linear combinations
\begin{eqnarray}
\sigma_A \equiv \frac{2 \sigma_X - \sigma_Y - \sigma_Z}{3} \;\;,\;\;
\sigma_E \equiv \frac{\sigma_Z - \sigma_Y}{\sqrt{3}} \;\;,\;\;
\sigma_T \equiv \frac{\sigma_X + \sigma_Y + \sigma_Z}{3} \;\;.
\end{eqnarray}
These combinations are also signal-orthogonal; moreover, in the case of equal arms the combination $T$ is insensitive to the signal, and it is often denoted as the null channel. We denote as $\Sigma_O$ the signal (measurement minus noise) in the $O$ channel (where $O$ is either A or E). The expectation value for the three-point function of the signal can formally be written as \cite{snr}
\begin{equation}
\left\langle \Sigma_O  \Sigma_{O'}  \Sigma_{O''}  \right\rangle = \sum_{\lambda_1,\lambda_2,\lambda_3}
\int \dd f_1 \dd f_2 \dd f_3 \, f_1 f_2 f_3 \; {\cal B}_{\lambda_1 \lambda_2 \lambda_3} \left( \vec{f}_1 ,  \vec{f}_2 ,
 \vec{f}_3  \right)  \; {\cal R}^{OO'O''}_{\lambda_1\lambda_2 \lambda_3} \left( \vec{f}_1 ,  \vec{f}_2 ,
 \vec{f}_3  \right) .
\end{equation}
\noindent
In this expression, $\lambda_i$ and $\vec{f}_i$ denote, respectively, the polarization and frequency (more precisely, the frequency vector, related to the wave vector by $\vec{f} = \vec{k}/2 \pi$) of the GWs involved in the correlator; ${\cal B}_{\lambda_i} ( \vec{f}_i)$ is the GW bispectrum, and  ${\cal R}$ the three-point response function. As the measurement is a time delay, $\Sigma_O$ has the dimension of an inverse mass. The bispectrum has mass dimension $-6$.
With these conventions, the response function is therefore dimensionless. Due to the planar nature of the instrument \cite{Smith:2016jqs}, the response function is invariant under parity, namely ${\cal R}_{\rm RRR} =  {\cal R}_{\rm LLL}  ,\; {\cal R}_{\rm RRL} =  {\cal R}_{\rm LLR} $ (and so on). Moreover, due to the highly symmetric configuration, only the EEE and AAE (and permutations) correlations of the channels are nonvanishing, and they are one the opposite of the other ${\cal R}^{\rm EEE} = - {\cal R}^{\rm AAE}$  \cite{snr}.
\begin{figure}[t!]
\includegraphics[width=.49\textwidth]{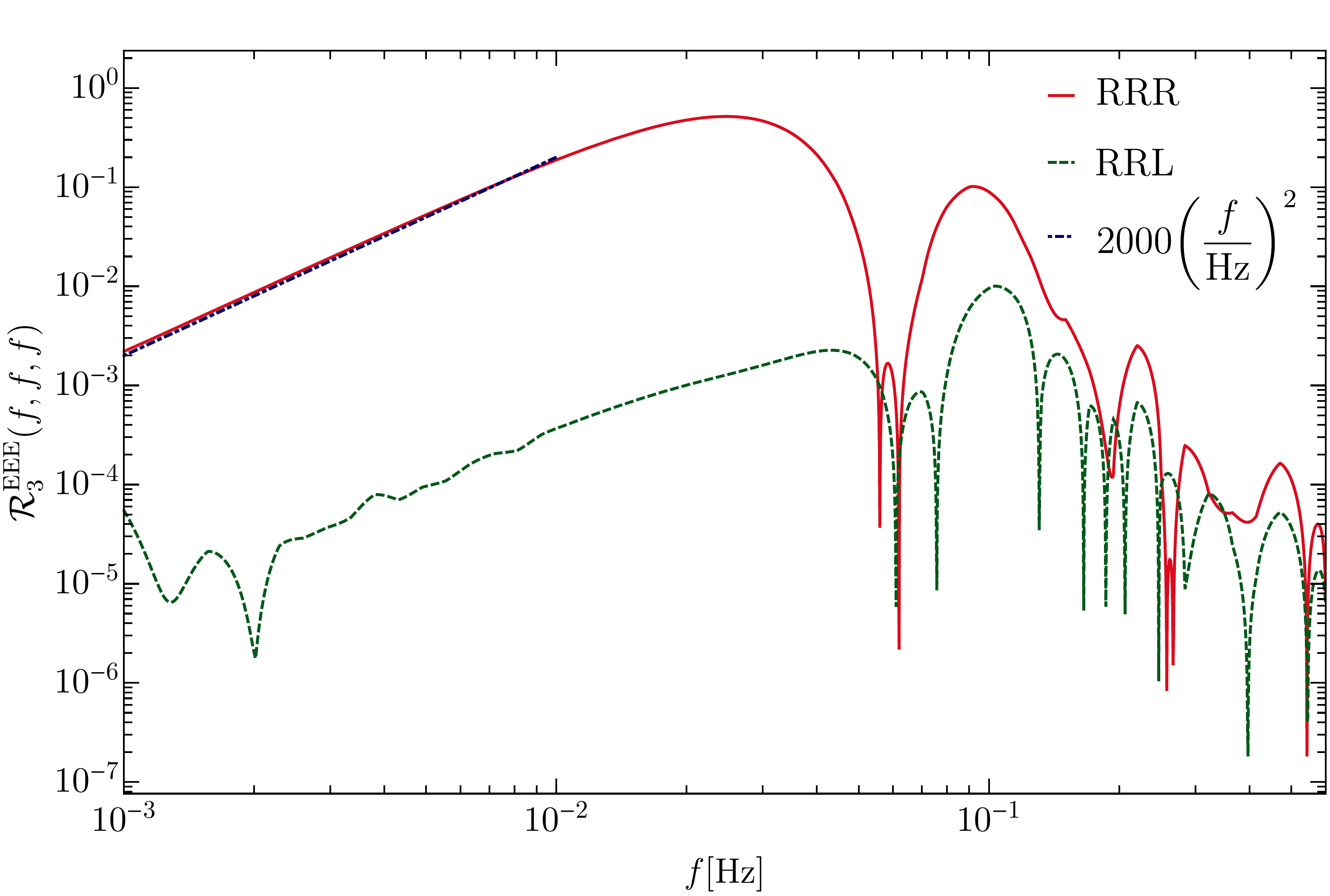}
\includegraphics[width=.49\textwidth]{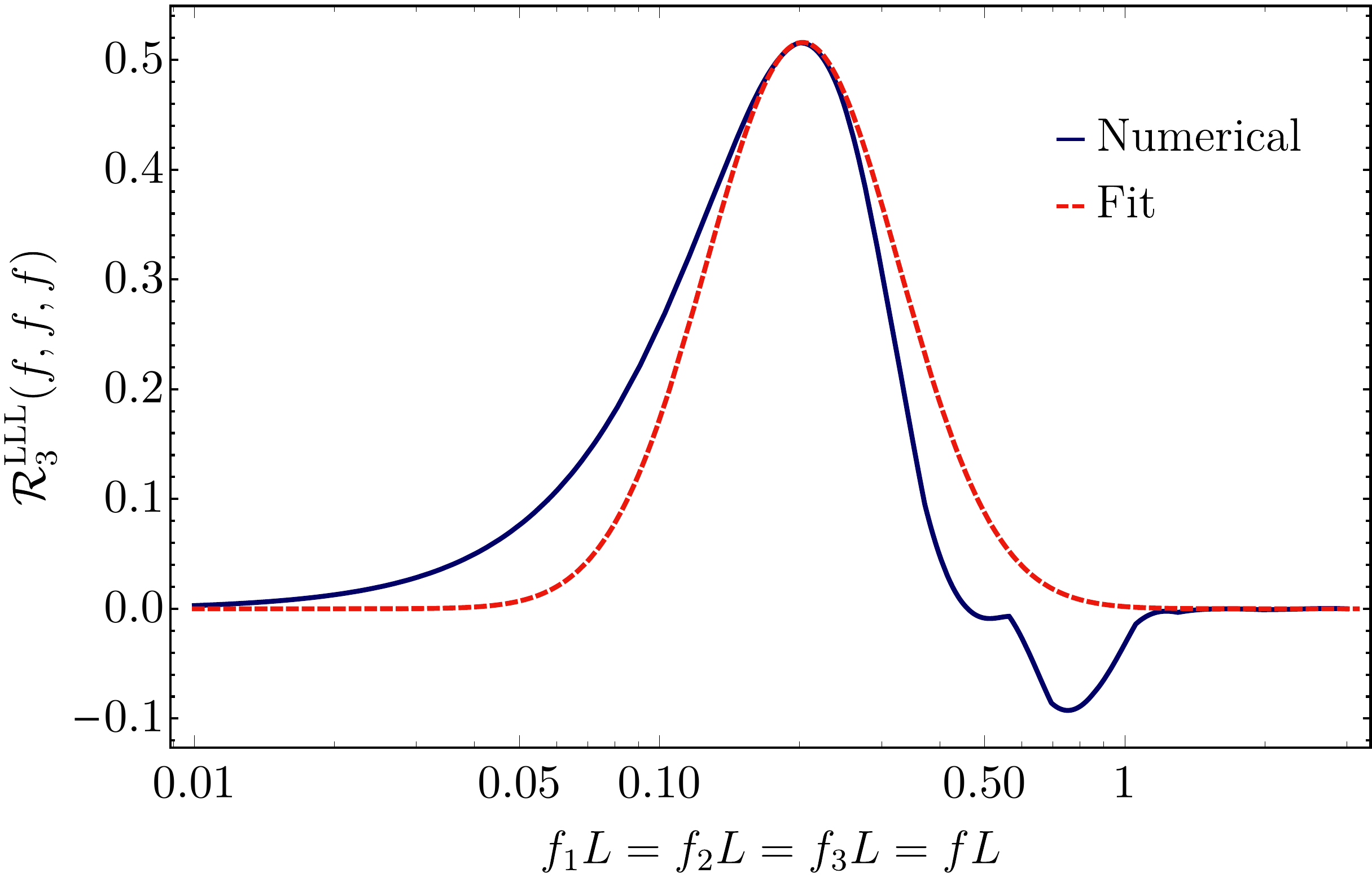}
\caption{{\bf Equilateral configuration.} {\it Left:}  Response function $\mathcal{R}_3^{\rm EEE}$ for RRR=LLL (red curve) and RRL=LLR (green curve),
as a function of the frequency. At $f=0.002~ {\rm Hz}$ the response function in the RRR case is about 500
times grater than in the RRL case. The blue dotted line indicates the expansion at small frequencies of Eq.~\eqref{R3-equi-fit}.
{\it Right:} Comparison between the response function on equilateral triangles and its 1d fit, see Eq.~\eqref{fit-R3-equil}.
}
\label{fig:R3-equil}
\end{figure}

All the relevant formulae for the computation of  ${\cal R}$  can be found in Section 3 of  Ref.~\cite{snr}, so we do not copy them here. In Figure \ref{fig:R3-equil} we show the two response functions ${\cal R}_{\rm LLL}^{\rm EEE}$ (red line) and ${\cal R}_{\rm LLR}^{\rm EEE}$ (green line) in the equilateral case $f_1=f_2=f_3$. We note that the instrument is significantly more sensitive to equal-helicity bispectrum LLL (and, equivalently, RRR). Moreover, we see that in the outmost left range shown in Figure \ref{fig:R3-equil} (left panel), the response function can be very well approximated by
\begin{equation}\label{R3-equi-fit}
{\cal R}_{\rm LLL}^{\rm EEE} \left( \vert f_i \vert \right) \simeq 2000 \, \left( \frac{f}{\rm Hz} \right)^2 ,\;\;\; 10^{-4} \la \frac{f}{\rm Hz} \la 10^{-2}.
\end{equation}
It is also useful to approximate the response function at its greatest peak. As shown in the right panel of Figure
\ref{fig:R3-equil}, the peak is well fitted by
\begin{equation}
{\cal R}_{\rm RRR}^{{\rm EEE},{\rm pk}} \left( f ,\, f  ,\, f \right) \simeq  \left[ A \, {\rm e}^{-\frac{ \ln^2 \frac{f}{f_r}}{2 \sigma^2}  } \right] \equiv \left[ {\cal R}_{\rm fit-1d} \left( f \right) \right]^3 ,
\label{fit-R3-equil}
\end{equation}
with (matching the peak position, amplitude, and curvature)
\begin{equation}
A \simeq 0.802 ,\;\; f_r \simeq \frac{0.203}{L} \simeq 0.0243 \, {\rm Hz}  ,\;\; \sigma^2 = 0.686 .
\end{equation}
In  Ref.  \cite{snr} only the equilateral $f_1 = f_2 = f_3$ and squeezed isosceles $f_1 = f_2 \gg f_3$ configurations were computed. Here, we compute and study the generic isosceles configuration $f_1 = f_2 \neq f_3$.
\begin{figure}[t!]
\includegraphics[width=.54\textwidth]{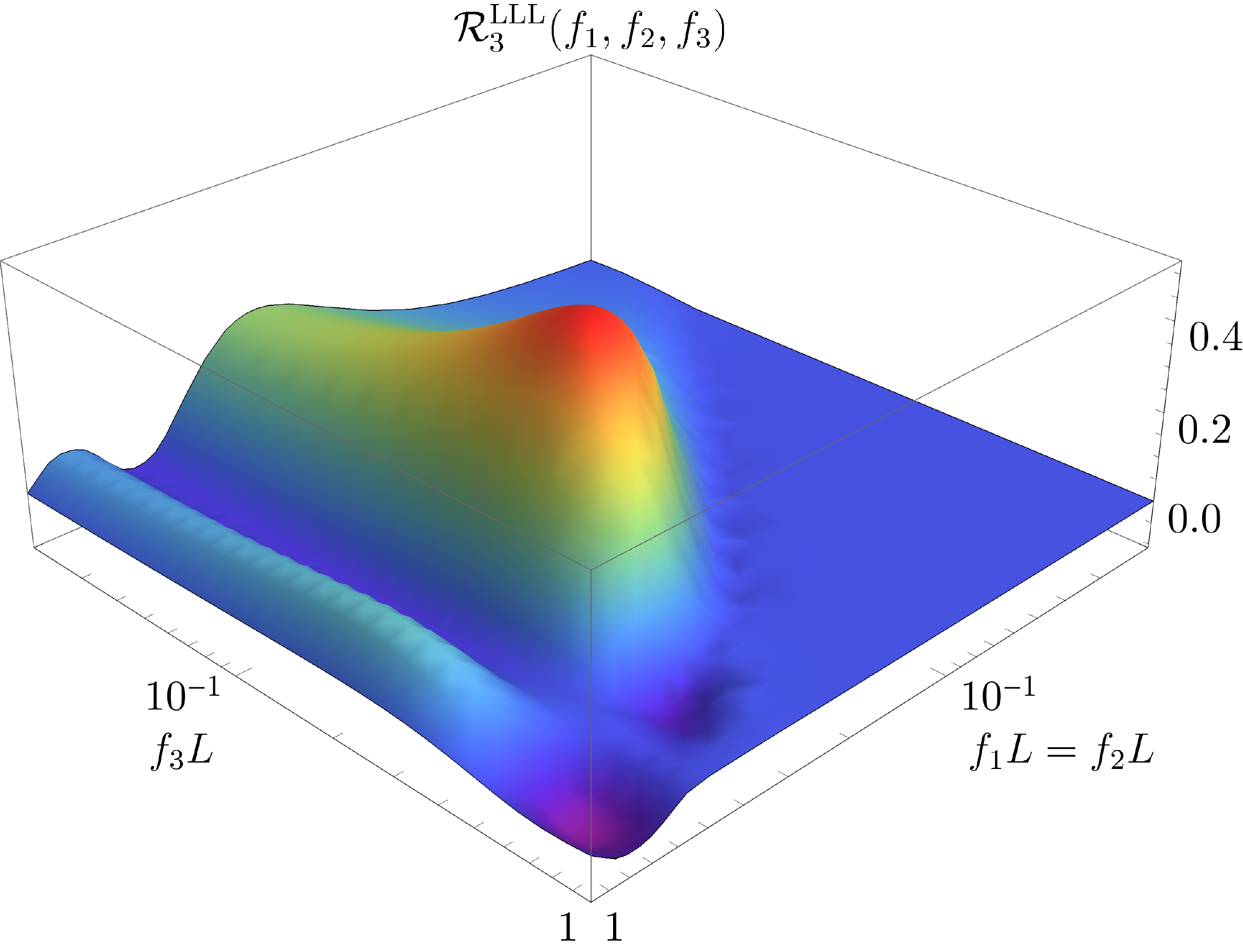}
\includegraphics[width=.44\textwidth]{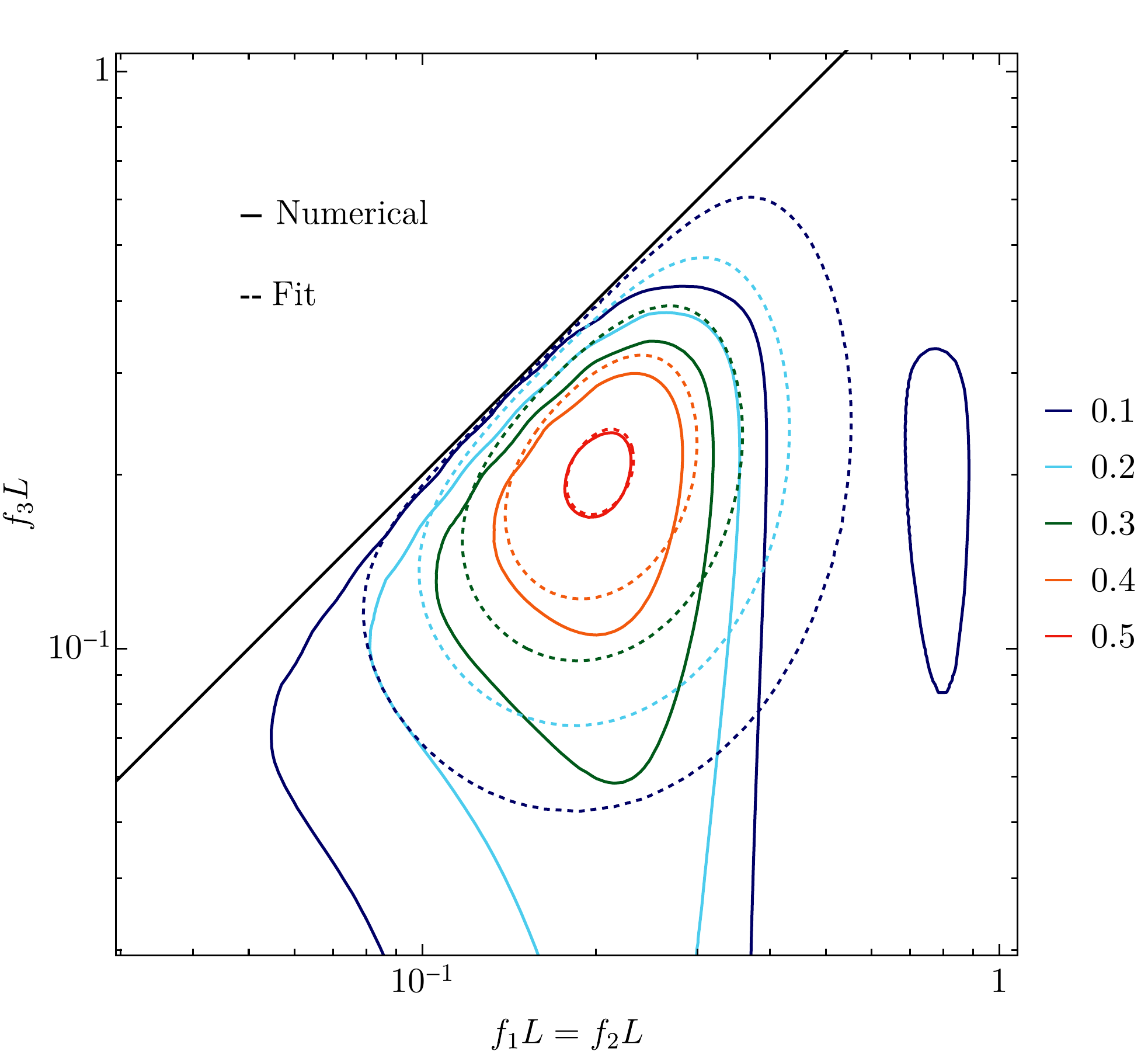}
\caption{{\bf Isosceles configuration.} {\it Left:} 3d plot of the response function $\mathcal{R}_3^{\rm LLL} (f_1,f_2,f_3)$ in the isosceles configuration. {\it Right:}  Contour plot of ${\cal R}_{\rm RRR}^{\rm EEE} \left( f_1 ,\, f_2  ,\, f_3 \right)$ for EEE and for RRR. We verified that the response function for AAE (and EAA) is the opposite to the one shown here. The black  solid line corresponds to the $f_3 = 2 f_1$ folded case (notice that the axes are in log units), and the upper-left corner does not exist (it violates the triangular inequality). The bottom part of the region corresponds to squeezed configurations. The dashed lines correspond to the 3d fit for the response function on isosceles triangles \eqref{3dfitR3}.}
\label{fig:R3-isosc}
\end{figure}

As we see in Fig. \ref{fig:R3-isosc} the response function is peaked in the equilateral configuration, at $f_1=f_2=f_3 = f_r$. A significant result is also found in the squeezed limit $f_3 \ll f_1 = f_2 \simeq f_r$.
The fitting formula (\ref{fit-R3-equil}) can be used to provide a fit of the generic shape next to the peak
\begin{equation} \label{3dfitR3}
{\cal R}_{\rm RRR}^{\rm EEE,{\rm pk}} \left( f_1 ,\, f_2  ,\, f_3 \right) \simeq {\bar A} \left( f_1 ,\, f_2 ,\, f_3 \right)  \cdot  {\cal R}_{\rm fit-1d} \left( f_1 \right) \,  {\cal R}_{\rm fit-1d} \left( f_2 \right) \,  {\cal R}_{\rm fit-1d} \left( f_3 \right) ,
\end{equation}
where
\begin{equation}
{\bar A} \left( f_1 ,\, f_2 ,\, f_3 \right) \equiv \frac{\sqrt{
\left( f_1 + f_2 + f_3 \right) \left( - f_1 + f_2 + f_3 \right)  \left( f_1 - f_2 + f_3 \right)  \left( f_1 + f_2 - f_3 \right) }}{\sqrt{3} \left( f_1 f_2 f_3 \right)^{2/3}}
\end{equation}
is the ratio between the area of a triangle of sides $f_1,f_2,f_3$, divided by the area of an equilateral triangle of sides $\left( f_1 f_2 f_3 \right)^{1/3}$. This factor evaluates to one at the peak, and it suppresses the bispectrum at its boundaries, so that the fitting formula (\ref{fit-R3-equil}) can be integrated over all possible shapes. In the right panel of the  Fig. \ref{fig:R3-isosc} the fitting function  (\ref{fit-R3-equil}) is compared against the exact bispectrum for isosceles shapes.


\bigskip

\begin{thebibliography}{99}
	
\bibitem{bertone} G.~Bertone and T.~M.~P.~Tait,
  Nature {\bf 562}, no. 7725, 51 (2018).



\bibitem{ligo} B. P. Abbott et al. [LIGO Scientific and Virgo Collaborations], Phys. Rev. Lett. 116, 061102 (2016)
\arXiv{1602.03837}{gr-qc}

\bibitem{Bird:2016dcv}
  S.~Bird, I.~Cholis, J.~B.~Muñoz, Y.~Ali-Haïmoud, M.~Kamionkowski, E.~D.~Kovetz, A.~Raccanelli and A.~G.~Riess,
  Phys.\ Rev.\ Lett.\  {\bf 116}, no. 20, 201301 (2016)
  \arXiv{1603.00464}{astro-ph.CO}.

\bibitem{Clesse:2016vqa}
  S.~Clesse and J.~García-Bellido,
  Phys.\ Dark Univ.\  {\bf 15}, 142 (2017)
  \arXiv{1603.05234}{astro-ph.CO}.

\bibitem{Sasaki:2016jop}
  M.~Sasaki, T.~Suyama, T.~Tanaka and S.~Yokoyama,
  Phys.\ Rev.\ Lett.\  {\bf 117}, no. 6, 061101 (2016)
  Erratum: [Phys.\ Rev.\ Lett.\  {\bf 121}, no. 5, 059901 (2018)]
  \arXiv{1603.08338}{astro-ph.CO}.

\bibitem{juan} J.~García-Bellido,
  J.\ Phys.\ Conf.\ Ser.\  {\bf 840}, no. 1, 012032 (2017)
  \arXiv{1702.08275}{astro-ph.CO}.

\bibitem{revPBH}
M.~Sasaki, T.~Suyama, T.~Tanaka and S.~Yokoyama,
Class.\ Quant.\ Grav.\  {\bf 35}, no. 6, 063001 (2018)
\arXiv{1801.05235}{astro-ph.CO}.

  \bibitem{revPBH1} L.~Barack {\it et al.},
  \arXiv{1806.05195}{gr-qc}.

\bibitem{s1} P.~Ivanov, P.~Naselsky and I.~Novikov,
Phys.\ Rev.\ D {\bf 50}, 7173 (1994).

\bibitem{s2} J.~Garc\'{\i}a-Bellido, A.D.~Linde and D.~Wands,
Phys.\ Rev.\ D {\bf 54} (1996) 6040
\arXivold{astro-ph/9605094}.

\bibitem{s3} 
P.~Ivanov, Phys.\ Rev.\ D {\bf 57}, 7145 (1998)
\arXivold{astro-ph/9708224}.

\bibitem{Espinosa:2017sgp}
  J.~R.~Espinosa, D.~Racco and A.~Riotto,
  Phys.\ Rev.\ Lett.\  {\bf 120} (2018) no.12,  121301
  \arXiv{1710.11196}{hep-ph}.

\bibitem{musco} I.~Musco,
\arXiv{1809.02127}{gr-qc}.


\bibitem{ng} G.~Franciolini, A.~Kehagias, S.~Matarrese and A.~Riotto,
JCAP {\bf 1803}, no. 03, 016 (2018)
\arXiv{1801.09415}{astro-ph.CO}.

\bibitem{Acquaviva:2002ud}
V.~Acquaviva, N.~Bartolo, S.~Matarrese and A.~Riotto,
Nucl.\ Phys.\ B {\bf 667} (2003) 119
\arXivold{astro-ph/0209156}.

\bibitem{Mollerach:2003nq}
S.~Mollerach, D.~Harari and S.~Matarrese,
Phys.\ Rev.\ D {\bf 69} (2004) 063002
\arXivold{astro-ph/0310711}.

\bibitem{Ananda:2006af}
K.~N.~Ananda, C.~Clarkson and D.~Wands,
Phys.\ Rev.\ D {\bf 75} (2007) 123518
\arXivold{gr-qc/0612013}.

\bibitem{Baumann:2007zm}
D.~Baumann, P.~J.~Steinhardt, K.~Takahashi and K.~Ichiki,
Phys.\ Rev.\ D {\bf 76} (2007) 084019
\arXivold{hep-th/0703290}.

\bibitem{jap} 
Prog.\ Theor.\ Phys.\  {\bf 123}, 867 (2010)
Erratum: [Prog.\ Theor.\ Phys.\  {\bf 126}, 351 (2011)]
\arXiv{0912.5317}{astro-ph.CO}.

\bibitem{Lisa} N.~Bartolo {\it et al.},
JCAP {\bf 1612} (2016) no. 12, 026
\arXiv{1610.06481}{astro-ph.CO}.

\bibitem{gbp} J.~Garcia-Bellido, M.~Peloso and C.~Unal,
JCAP {\bf 1709}, no. 09, 013 (2017)
\arXiv{1707.02441}{astro-ph.CO}.


\bibitem{k} A.~Katz, J.~Kopp, S.~Sibiryakov and W.~Xue,
\arXiv{1807.11495}{astro-ph.CO}.

\bibitem{japanese} K.~Inomata, M.~Kawasaki, K.~Mukaida, Y.~Tada and T.~T.~Yanagida,
Phys.\ Rev.\ D {\bf 96} (2017)  043504
\arXiv{1701.02544}{astro-ph.CO}.

\bibitem{hsc} H.~Niikura {\it et al.},
\arXiv{1701.02151}{astro-ph.CO}.

\bibitem{hsc1} K.~Inomata, M.~Kawasaki, K.~Mukaida and T.~T.~Yanagida,
Phys.\ Rev.\ D {\bf 97}, no. 4, 043514 (2018)
\arXiv{1711.06129}{astro-ph.CO}.

\bibitem{neut} F.~Capela, M.~Pshirkov and P.~Tinyakov,
Phys.\ Rev.\ D {\bf 87}, no. 12, 123524 (2013)
\arXiv{1301.4984}{astro-ph.CO}.

\bibitem{Brandt:2016aco}
  T.~D.~Brandt,
  Astrophys.\ J.\  {\bf 824}, no. 2, L31 (2016)
  \arXiv{1605.03665}{astro-ph.GA}.

\bibitem{Li:2016utv}
  T.~S.~Li {\it et al.} [DES Collaboration],
  Astrophys.\ J.\  {\bf 838} (2017) no.1,  8
  \arXiv{1611.05052}{astro-ph.GA}.

\bibitem{Zumalacarregui:2017qqd}
  M.~Zumalacarregui and U.~Seljak,
  Phys.\ Rev.\ Lett.\  {\bf 121} (2018) no.14,  141101
  \arXiv{1712.02240}{astro-ph.CO}.

\bibitem{Garcia-Bellido:2017imq}
  J.~Garcia-Bellido, S.~Clesse and P.~Fleury,
  Phys.\ Dark Univ.\  {\bf 20} (2018) 95
  \arXiv{1712.06574}{astro-ph.CO}.

\bibitem{disk-acc}
  V.~Poulin, P.~D.~Serpico, F.~Calore, S.~Clesse and K.~Kohri,
  \arXiv{1707.04206}{astro-ph.CO}.

\bibitem{murgia} 
  R.~Murgia, G.~Scelfo, M.~Viel and A.~Raccanelli,
  \arXiv{1903.10509}{astro-ph.CO}.
  
\bibitem{cirelli} 
  M.~Boudaud and M.~Cirelli,
  Phys.\ Rev.\ Lett.\  {\bf 122}, no. 4, 041104 (2019)
  \arXiv{1807.03075}{astro-ph.HE}.



\bibitem{short} 
  N.~Bartolo, V.~De Luca, G.~Franciolini, A.~Lewis, M.~Peloso and A.~Riotto,
  Phys.\ Rev.\ Lett.\  {\bf 122}, no. 21, 211301 (2019)
 \arXiv{1810.12218}{astro-ph.CO}.


\bibitem{we} S.~Weinberg,
  Phys.\ Rev.\ D {\bf 69}, 023503 (2004)
  \arXivold{astro-ph/0306304}.

\bibitem{lrreview}   D.H.~Lyth and A.~Riotto,
Phys.\ Rept.\  {\bf 314} (1999) 1
\arXivold{hep-ph/9807278}.


\bibitem{errgw} J.~R.~Espinosa, D.~Racco and A.~Riotto,
JCAP {\bf 1809}, no. 09, 012 (2018)
\arXiv{1804.07732}{hep-ph}.

\bibitem{Kohri:2018awv}
K.~Kohri and T.~Terada,
 Phys.\ Rev.\ D {\bf 97} (2018) no.12,  123532
 \arXiv{1804.08577}{gr-qc}.

\bibitem{saito} R.~Saito and J.~Yokoyama,
Prog.\ Theor.\ Phys.\  {\bf 123}, 867 (2010)
Erratum: [Prog.\ Theor.\ Phys.\  {\bf 126}, 351 (2011)]
\arXiv{0912.5317}{astro-ph.CO}.


\bibitem{e} E.~Bugaev and P.~Klimai,
  Phys.\ Rev.\ D {\bf 81}, 023517 (2010)
  \arXiv{0908.0664}{astro-ph.CO}.

  \bibitem{Audley:2017drz}
H.~Audley {\it et al.},
\arXiv{1702.00786}{astro-ph.IM}.

\bibitem{Caprini:2015zlo}
C.~Caprini {\it et al.},
JCAP {\bf 1604} (2016) no.04,  001
\arXiv{1512.06239}{astro-ph.CO}.

\bibitem{snr}
N.~Bartolo {\it et al.},
\arXiv{1806.02819}{astro-ph.CO}.

\bibitem{Matsubara}
  T.~Matsubara,
  Astrophys.\ J.\ Suppl.\  {\bf 170}, 1 (2007)
  \arXivold{astro-ph/0610536}.


    \bibitem{hu} W.~Hu and A.~Cooray,
  Phys.\ Rev.\ D {\bf 63}, 023504 (2001)
  \arXivold{astro-ph/0008001}.

  \bibitem{vg} D.~S.~Gorbunov and V.~A.~Rubakov,
  ``Introduction to the theory of the early universe: Cosmological perturbations and inflationary theory,''
  Hackensack, USA: World Scientific (2011)
  
  \bibitem{sasakinew} R.~g.~Cai, S.~Pi and M.~Sasaki,
  \arXiv{1810.11000}{astro-ph.CO}.
  
\bibitem{compagno} 
  T.~Nakama, J.~Silk and M.~Kamionkowski,
  Phys.\ Rev.\ D {\bf 95}, no. 4, 043511 (2017)
  \arXiv{1612.06264}{astro-ph.CO}.
  
  \bibitem{Graham:2015apa}
  P.~W.~Graham, S.~Rajendran and J.~Varela,
  Phys.\ Rev.\ D {\bf 92} (2015) no.6,  063007
  \arXiv{1505.04444}{hep-ph}.

\bibitem{Gould:1991td}
  A.~Gould,
  Submitted to: Astrophys.~J.~Lett. (1991).

\bibitem{Nakamura:1997sw}
  T.~T.~Nakamura,
  Phys.\ Rev.\ Lett.\  {\bf 80} (1998) 1138.


\bibitem{corn} M.~R.~Adams and N.~J.~Cornish,
Phys.\ Rev.\ D {\bf 82}, 022002 (2010)
\arXiv{1002.1291}{gr-qc}.


\bibitem{Smith:2016jqs}
  T.~L.~Smith and R.~Caldwell,
  Phys.\ Rev.\ D {\bf 95} (2017) no.4,  044036
  \arXiv{1609.05901}{gr-qc}.



\end{thebibliography}
\end{document}